\documentclass[a4paper]{article}

\usepackage{latexsym}
\usepackage{amssymb}
\usepackage{array}

 \def\newplaintheorem#1#2#3{%
\newtheorem{#1plain}[#3]{#2}
\newenvironment{#1}{\begin{#1plain}\rm}{\end{#1plain}}}

\newplaintheorem{myexample}{Example}{theo}
\newplaintheorem{rmdefinition}{Definition}{theo}
\newplaintheorem{rmcoro}{Corollary}{theo}
\newplaintheorem{rmtheorem}{Theorem}{theo}
\newplaintheorem{rmlemma}{Lemma}{theo}
\newplaintheorem{propo}{Proposition}{theo}
\newplaintheorem{remark}{Remark}{theo}

\newtheorem{factplain}{Fact}
\newenvironment{fact}{\begin{factplain}\rm}{\end{factplain}}

\newcounter{partcounter}
\newcounter{subpartcounter}[partcounter]
\newcounter{casecounter}
\newcounter{subcasecounter}[casecounter]

\newenvironment{proof}
 {{\normalfont \vspace{1ex} \noindent {\scshape Proof.}}%
   \setlength{\parindent}{0pt}%
   \setlength{\parskip}{1.5ex plus 0.5ex minus 1.0ex}%
   \setcounter{partcounter}{1}
   \setcounter{casecounter}{1}}
 {\hspace*{1em}\hfill$\Box$}

\newenvironment{prooffact}{\noindent {\it Proof of Fact.}}%
{\qed}

\newcommand{\ol}[1]{{{\bf #1}}}

\newenvironment{program}{\tt \begin{tabbing}pro\= {\tt pro}\= clause
    \kill}{\end{tabbing}}

\newcommand{\la}{\ensuremath{\:\leftarrow\:}}

\newcommand{\In}{{\it In}}

\newcommand{\Var}{{\it Var}}
\newcommand{\Dom}{{\it Dom}}
\newcommand{\Ran}{{\it Ran}}

\newcommand{\U}{{\ensuremath{\:\cup\:}}}

\newcommand{\qed}{\hspace*{0.5em} \hfill $\Box$}

\newcommand{\Dersym}{\ensuremath{\longrightarrow}}

\def\parentarabici{\renewcommand{\theenumi}{\arabic{enumi}}
\renewcommand{\labelenumi}{(\theenumi)}}

\def\parentalphi{\renewcommand{\theenumi}{\alph{enumi}}
\renewcommand{\labelenumi}{(\theenumi)}}

\begin{document}


\title{Semantics and Termination of Simply-Moded Logic Programs with Dynamic
  Scheduling\thanks{This paper is the long version
of~\cite{BERS01}. It contains the proofs omitted there for space reasons.}}

\author{
\hspace*{3.8em}
Annalisa Bossi\thanks{
Universit\`a di  Venezia,
  \texttt{\{bossi,srossi\}@dsi.unive.it}}
\hspace*{0.7em}
\and 
Sandro Etalle\thanks{
Universiteit Maastricht, 
\texttt{etalle@cs.unimaas.nl}}
\thanks{
CWI, Amsterdam, 
\texttt{jan.smaus@cwi.nl}}
\hspace*{4.5em}
\and 
\addtocounter{footnote}{-3}
Sabina Rossi\footnotemark 
\addtocounter{footnote}{1}
\and Jan-Georg Smaus\footnotemark 
}
\date{\vspace*{-3ex}}
\maketitle

\begin{abstract}
In logic programming, {\em dynamic scheduling} refers to
a situation where the selection of the atom in each resolution (computation)
step is determined at runtime, as opposed to a fixed selection rule
such as the left-to-right one of Prolog.  This has
applications e.g.\ in parallel programming. A mechanism to control dynamic scheduling
is provided in existing languages in the form of 
{\em delay declarations}.  

Input-consuming derivations were introduced
to describe
dynamic sched\-uling while abstracting from the technical details.  In
this paper, we first formalise the relationship between delay
declarations and input-consuming derivations, showing in many
cases a one-to-one correspondence. Then, we define a
model-theoretic semantics for input-consuming derivations of
simply-moded programs. Finally, for this class of
programs, we provide a necessary and sufficient criterion for
termination.
\end{abstract}

\section{Introduction}

\subsection{Background}
Logic programming is based on giving a computational
interpretation to a fragment of first order logic.
 Kowalski\index{Kowalski}~\cite{K79}
advocates the separation of the {\em logic}\index{logic} and 
{\em control}\index{control}  aspects of
a logic program and has coined the famous formula
\begin{quote}
Algorithm = Logic + Control.\index{Algorithm = Logic + Control}
\end{quote}
The programmer should be responsible for the logic part. The control
should be taken care of by the logic programming system.

In reality, logic programming is far from this ideal. Without the
programmer being aware of the control and writing programs
accordingly, logic programs would usually be hopelessly inefficient
or even non-terminating. 

One aspect of control in logic programs is the {\em selection rule},
stating which atom in a query is selected in each derivation step.
The standard selection rule in logic programming languages is the
fixed left-to-right rule of Prolog. While this rule provides
appropriate control for many applications, there are situations, e.g.\ 
in the context of parallel execution or the test-and-generate
paradigm, that require a more flexible control mechanism, namely,
\emph{dynamic scheduling}, where the selectable atoms are determined
at runtime.  Such a mechanism is provided in modern logic programming
languages in the form of \emph{delay declarations} \cite{Na86}.

To demonstrate that on the one hand, the left-to-right selection rule
is sometimes inappropriate, but that on the other hand, the selection
 mechanism must be controlled in some way, consider the following 
programs \texttt{APPEND} and \texttt{IN\_ORDER}
\begin{program}
\% \> append(Xs,Ys,Zs) \la Zs \textrm{is the result of concatenating the lists} Xs \textrm{and} Ys \\
\> append([H|Xs],Ys,[H|Zs]) \la \\
\> \> append(Xs,Ys,Zs).\\
\> append([],Ys,Ys).\\[2mm]
\% \> in\_order(Tree,List) \la List \textrm{is an ordered list of the nodes of} Tree\\
\> in\_order(tree(Label,Left,Right),Xs) \la \\
\> \> in\_order(Left,Ls),\\
\> \> in\_order(Right,Rs),\\ 
\> \> append(Ls,[Label|Rs],Xs).\\
\> in\_order(void,[]).
\end{program}
together with the query  (\texttt{read\_tree} and \texttt{write\_list} are defined
elsewhere)
\[
\texttt{q}: \texttt{read\_tree(Tree), in\_order(Tree,List), write\_list(List)}.
\]
If \texttt{read\_tree} cannot read the
whole tree at once 
--- say, it receives the input from a stream ---
it would be nice to be able to run the ``processes'' \texttt{in\_order} and
\texttt{write\_list} on the available input.  This can only be done if
one uses a dynamic selection rule (Prolog's rule would call
\texttt{in\_order} only after \texttt{read\_tree} has finished, while
other fixed rules would immediately diverge).  In order to avoid
nontermination one should adopt appropriate delay declarations,
namely
\begin{program}
\> delay in\_order(T,\_) until nonvar(T).\\
\> delay append(Ls,\_,\_) until nonvar(Ls).\\
\> delay write\_list(Ls,\_) until nonvar(Ls).
\end{program}
These declarations avoid that \texttt{in\_order}, \texttt{append}
and \texttt{write\_list} are selected ``too early'',
i.e.\ when their
arguments are not ``sufficiently instantiated''.
Note that instead of having interleaving ``processes'', one can also
select several atoms in {\em parallel}, as long as the delay
declarations are respected. 
This approach to parallelism has been first proposed in \cite{Nai88}
  and 
 ``has an important advantage over the
 ones proposed in the literature in that it allows us to parallelise
 programs written in a large subset of Prolog by merely adding to them
 delay declarations, so \emph{without modifying} the original
 program'' \cite{AL95}.

Compared to other mechanisms for user-defined control, e.g., using the 
cut operator in connection with built-in predicates that test for the
instantiation of a variable (\verb:var: or \verb:ground:), delay
declarations are more compatible with the declarative character of
logic programming. Nevertheless, many important declarative properties that have been
proven for logic programs do not apply to programs
with delay declarations. The problem is mainly related to 
{\em deadlock}.

Essentially, for such programs the well-known equivalence
between model-theoretic and operational semantics does not hold.  For
example, the query \texttt{append(X,Y,Z)} does not succeed (it
\emph{deadlocks}) and this is in contrast with the fact that
(infinitely many) instances of \texttt{append(X,Y,Z)} are contained in
the least Herbrand model of \texttt{APPEND}.  This shows that a
model-theoretic semantics in the classical sense is not achievable, in
fact the problem of finding a suitable declarative semantics is still
open.  Moreover, while for the left-to-right
selection rule there are results that allow us to characterise when a
program is terminating, these results do not apply any longer in
presence of dynamic scheduling.

\subsection{Contributions}
This paper contains essentially four contributions
tackling the above problems.

In order to provide a characterisation of dynamic scheduling that is
reasonably abstract and hence amenable to semantic analysis, 
we consider \emph{input-consuming derivations}
\cite{S99}, a formalism similar to \emph{Moded GHC}
\cite{UM94}. In an input-consuming derivation,
only atoms whose input arguments are not instantiated through the
unification step may be selected.
Moreover, we restrict our attention to the class of \emph{simply-moded} programs, which
are programs that are, in a well-defined sense,
consistent wrt.~the modes. As also shown by the
 benchmarks in Sec.\ \ref{app:benchmarks}, most practical
programs are simply-moded.  
We analyse the relations between input-consuming derivations
 and programs with delay declarations. We demonstrate that
  under some statically verifiable conditions, 
  input-consuming derivations are exactly the ones satisfying the
  (natural) delay declarations of  programs.

We define a denotational semantics which enjoys a
model-theoretical reading and has a bottom-up constructive definition.
We show that it is compositional, correct and fully abstract
wrt.~the computed answer substitutions of successful derivations.
E.g., it captures the fact that the query
\texttt{append(X,Y,Z)} does not succeed.

Since dynamic scheduling also allows for parallelism, it is sometimes
important to model the result of \emph{partial} (i.e., incomplete)
derivations.  For instance, one might have queries (processes) that
never terminate, which by definition may never reach the state of
\emph{success}, i.e.\ of successful completion of the computation.
Therefore, we define a second semantics which enjoys the same properties as
  the one above. We demonstrate that it
  is correct, fully abstract and compositional wrt.~the computed substitutions of
  partial  derivations.
  We then have a uniform (in our opinion elegant) framework 
  allowing us to model both successful and partial computations.
  
Finally, we study the problem of termination of input-consuming
  programs. We present a result which characterises termination
  of simply-moded input-consuming programs. This result is based on 
the semantics mentioned in the previous paragraph.

The rest of this paper is organised as follows. 
The next section introduces some preliminaries.
Section~\ref{sec:IC-prop} shows properties of input-consuming derivations
that are needed in the proofs.
Section~\ref{sec:IC-and-delay} defines delay
declarations, and formally compares them to input-consuming derivations.
Section~\ref{sec:semantics} provides a result on denotational
semantics for input-consuming derivations, first for complete
derivations, then for incomplete
(input-consuming) derivations. 
Section~\ref{sec:termination} provides a sufficient and necessary
criterion for termination of programs using input-consuming derivations.
Section~\ref{app:benchmarks} surveys some benchmark programs.
Section~\ref{sec:comparison} concludes.

\section{Preliminaries}
\label{sec:prel}

The reader is assumed to be familiar with the terminology and the
basic results of the semantics of logic programs
\cite{Apt90,Apt97,Llo87}. Following \cite{Apt97}, we use boldface
characters to denote sequences of objects: $\ol t$ denotes a sequence
of terms, $\ol B$ is a query (i.e., a possibly empty sequence
of atoms).  The empty query is denoted by $\Box$.
The relation symbol of an atom $A$ is denoted $Rel(A)$.
The set of variables occurring in a syntactic object $o$ is denoted
${\it Var}(o)$. We say that $o$ is \emph{linear} if every variable
occurs in it at most once.  Given a \emph{substitution} $\sigma =
\{x_1/t_1 ,\ldots ,x_n/t_n\}$, we say that $\{x_1,\ldots,x_n\}$ is its
\emph{domain} (denoted by $\Dom(\sigma)$), and $\Var(\{t_1
,\ldots,t_n\})$ is its \emph{range} (denoted by $\Ran(\sigma)$).  Note
that $\Var(\sigma) = \Dom(\sigma) \U \Ran(\sigma)$.
If $t_1 ,\ldots,t_n$ is a
permutation of $x_1 ,\ldots,x_n$ then we say that $\sigma$ is a
\emph{renaming}.  
The \emph{composition} of substitutions is denoted by juxtaposition
($x\theta \sigma= (x\theta)\sigma$).  We say that a term $t$ is an
\emph{instance} of $t'$ iff for some $\sigma$, $t = t'\sigma$;
further, $t$ is a \emph{variant} of $t'$, written $t\approx t'$, iff
$t$ and $t'$ are instances of each other.  A substitution $\theta$ is
a \emph{unifier} of terms $t$ and $t'$ iff $t\theta=t'\theta$. We
denote by $\emph{mgu}(t,t')$ any \emph{most general unifier}
(\emph{mgu}, in short) of $t$ and $t'$.  A query $Q: {\ol A},B,{\ol
  C}$ and a clause $c: H\leftarrow {\ol B}$ (variable disjoint with
$Q$) yield the resolvent $({\ol A},{\ol B},{\ol C})\theta$ with
$\theta = mgu(B,H)$.  We say that ${\ol A},B,{\ol C}
\stackrel{\theta}\Longrightarrow ({\ol A},{\ol B},{\bf C})\theta$ is a
\emph{derivation step (using $c$)}, and call $B$ the \emph{selected
  atom}.  A \emph{derivation of} $P \U \{ Q \}$ is a sequence of
derivation steps $Q\stackrel{\theta_1}\Longrightarrow
Q_1\stackrel{\theta_2}\Longrightarrow\cdots$ using (variants of)
clauses in the program $P$.  A finite derivation
$Q\stackrel{\theta_1}\Longrightarrow\cdots
\stackrel{\theta_n}\Longrightarrow Q_n$ is also denoted
$Q\stackrel{\vartheta} \Dersym_P Q_n$, where
$\vartheta=\theta_1\dots\theta_n$.  The restriction of $\vartheta$ to
$Q$ is a \emph{computed answer substitution} (\emph{c.a.s.}). If
$Q_n=\Box$, the derivation is \emph{successful}.

\subsection{Delay Declarations}
Logic programs with delay declarations 
consist of two parts: a set of clauses and a set of delay
declarations, one for each of its predicate symbols. A 
{\em delay declaration} associated with an $n$-ary predicate symbol $p$
has the form
$$
{\tt delay} \;\;\; p(t_1,\ldots,t_n)
\;\;\; {\tt until} \;\;\; {\it Cond}(t_1,\ldots,t_n)
$$ 
where ${\it Cond}(t_1,\ldots,t_n)$ is a formula in some assertion 
language~\cite{HL94}. 
A derivation is \emph{delay-respecting} if an atom 
$p(t_1,\ldots,t_n)$ is selected only if 
${\it Cond}(t_1,\ldots,t_n)$
is satisfied.  In this case, we
also say that the atom $p(t_1,\ldots,t_n)$ is {\em selectable}.
In particu\-lar, we consider delay declarations of the
form
$$
{\tt delay} \;\;\; p({\tt X}_1,\ldots,{\tt X}_n)
\;\;\; {\tt until \;\;\; nonvar}({\tt X}_{i_1})\wedge\dots\wedge 
  {\tt nonvar}({\tt X}_{i_k}).
  $$
  where $1\leq i_1< \dots < i_k \leq n$.\footnote{For the case that
    $k=0$, the empty conjunction might be denoted as $\tt true$, or
    the delay declaration might simply be omitted.}  
  The condition 
${\tt nonvar}(t_{i_1})\wedge\dots\wedge {\tt nonvar}(t_{i_k})$ 
is satisfied if and only if
  $t_{i_1},\dots,t_{i_k}$ are non-variable terms.  
Such delay declarations
  are equivalent to the {\tt block} declarations of SICStus Prolog~\cite{sicstus98}.

\subsection{Moded Programs}
A \emph{mode} indicates how a predicate should be used.

\begin{rmdefinition}
  A \emph{mode} for a predicate symbol $p$ of arity $n$, is a function
  $m_p$ from $\{1,\ldots,n\}$ to $\{{\it In},{\it Out}\}$.  
\qed
\end{rmdefinition}

If $m_p(i)={\it In}$ (resp.\emph{ Out}), we say that $i$ is an ${\it
  input}$ (resp.\emph{ output}) \emph{position of} $p$.
We denote by ${\it In}(Q)$ (resp.${\it Out}(Q)$) the sequence of terms
filling in the input (resp.\ output) positions of predicates in $Q$.
Moreover, when writing an atom as $p({\ol s},{\ol t})$, we are
indicating that ${\ol s}$ is the sequence of terms filling in its input
positions and ${\ol t}$ is the sequence of terms filling in its
output positions.

%
%
The notion of simply-moded program is due to Apt and Etalle~\cite{AE93}.

\begin{rmdefinition}
\label{simply-moded-def}
%
A clause 
$p({\ol t}_0,{\ol s}_{n+1})\la 
p_1({\ol s}_1,{\ol t}_1),\ldots,p_n({\ol s}_n,{\ol t}_n)$
is 
  \emph{simply-moded} iff ${\ol t}_1,\ldots,{\ol t_n}$ is a linear
  vector of variables and for all $i\in [1,n]$
 
  $${\it Var}({\ol t}_i)\cap {\it Var}({\ol t}_0) = \emptyset
\quad \mbox{and} \quad
{\it Var}({\ol t}_i)\cap \bigcup_{j=1}^i{\it Var}({\ol
    s}_j)=\emptyset.$$
  
A query $\ol B$ is  \emph{simply-moded} iff the clause $q
  \la \ol B$ is simply-moded, where $q$ is any variable-free atom.
A program is simply-moded iff all of its clauses are.
\qed
\end{rmdefinition}

Thus, a clause is simply-moded if the output positions of body atoms
are filled in by distinct variables, and every variable occurring in
an output position of a body atom does not occur in an earlier input
position.  In particular, every unit clause is simply-moded.
Notice also that programs \texttt{APPEND} and \texttt{IN\_ORDER} are
  simply-moded wrt.\ the modes
\texttt{append(In,In,Out)} and
\texttt{in\_order(In,Out)}. 

\subsection{Input-Consuming Derivations}\label{subsec:IC}
Input-consuming derivations are a
formalism for describing dynamic scheduling in an abstract
way \cite{S99}. 

\begin{rmdefinition}
A derivation step ${\ol A},B,{\ol C}
  \stackrel{\theta}\Longrightarrow $
$({\ol A},{\ol B},{\ol C})\theta$ 
is \emph{input-consuming} if and only if 
$\In(B)\theta = \In(B)$.
A derivation is \emph{input-consuming} iff all its
derivation steps are input-consuming. \qed
\end{rmdefinition}

Thus, allowing only input-consuming derivations is a form of dynamic
sched\-uling, since whether or not an atom can be selected 
depends on its degree of instantiation 
at runtime.
If no atom is resolvable via an input-consuming derivation step, the
query \emph{deadlocks}.\footnote{
Notice that  there is a difference between this notion of  deadlock
and the one used for programs with delay declarations;
see~\cite{BER99} for a detailed discussion.}

It has been shown that the input-consuming\footnote{The notion of 
{\em input-consuming} is not used, but it is said that the input of
the selected atom must be an instance of the input of the head, which is
in fact a necessary condition for a derivation step to be input-consuming.}
 resolvent of a simply-moded
query using a simply-moded clause is simply-moded \cite[Lemma~30]{AL95}.

\section{Input-Consuming Derivations and Delay Declarations}
\label{sec:IC-and-delay}

In this section, we show a correspondence between input-consuming
derivations and delay declarations.

\begin{myexample}\mbox{ }
\label{exa:i-c}
Consider again the delay declaration 
\begin{program}
\> delay append(Ls, \_, \_) until nonvar(Ls).
\end{program}
It is easy to check that \emph{every} derivation starting in a query
\texttt{append($t$,$s$,X)}, where \texttt{X} is a variable disjoint
from $s$ and $t$, is input-consuming 
wrt.\ \texttt{append(In,In,Out)}
{\em iff} it respects the delay
declaration.  
\end{myexample}

To show the correspondence between delay declarations and
input-consuming derivations suggested by this example, we need
some further definitions.  We call a term $t$ \emph{flat} if $t$ has
the form $f(x_1,\ldots,x_n)$ where the $x_i$ are distinct variables.
Note that constants are flat terms.  The significance of flat term
arises from the following observation: if $s$ and $t$ are unifiable,
$s$ is non-variable and $t$ is flat, then $s$ is an instance of $t$.
Think here of $s$ being a term in an input position of a selected
atom, and $t$ being the term in that position of a clause head.

\begin{rmdefinition} A program $P$ is  \emph{input-consistent} 
  iff for each clause $H \la \ol B$ of it, the family of terms filling
  in the input positions of $H$ is linear, and consists of variables and
  flat terms.  \qed
\end{rmdefinition}

We also consider here delay declarations   of a
restricted type.
\begin{rmdefinition}\label{def:simple}
A program with delay declarations is 
  \emph{simple} if every delay declaration is of the form
$$
{\tt delay} \;\;\; p({\tt X}_1,\ldots,{\tt X}_n)
\;\;\; {\tt until \;\;\; nonvar}({\tt X}_{i_1})\wedge\dots\wedge 
  {\tt nonvar}({\tt X}_{i_k}).
  $$
where $i_1,\dots,i_k$ is a subset of the input positions of $p$.

Moreover, we say that the positions $i_1,\dots,i_k$ of $p$ are
\emph{controlled}, while the other input positions of $p$ are 
\emph{free}.
\qed
\end{rmdefinition}

Thus the controlled positions are those ``guarded'' by a delay
declaration. The main result of this section shows that,
 under some circumstances, using delay declarations is equivalent
to restricting to input-consuming derivations.
The first statement of the theorem has been shown previously 
\cite[Theorem 7.9]{Smaus-thesis}, although not exactly for the same
class of programs. We prefer to give a proof here.

\begin{rmlemma}
\label{lem:delay=IC} 
Let $P$ be simply-moded,
  input-consistent and simple. Let $Q$ be a simply-moded query.
\begin{itemize}
\item If for every clause $H \la \ol B$ of $P$, $H$ contains variables
  in its free positions, then every derivation of $P\cup \{Q\}$
  respecting the delay declarations is  input-consuming
  (modulo renaming).
\item If in addition for every clause $H \la \ol B$ of $P$, the head $H$
  contains flat terms in its controlled positions, then every
  input-consuming derivation of $P\cup \{Q\}$ respects the delay
  declarations. 
\end{itemize}

\begin{proof}
It is sufficient to show the result for the first step. The general
result follows from the persistence of simply-modedness under
input-consuming derivation steps  \cite[Lemma~30]{AL95}.
Let $A = p({\ol s},{\ol t})$ be the atom in $Q$ selected in the
step and $H = p({\ol v},{\ol u})$.

We prove the first statement.  Clearly $A$ and $H$ are
unifiable. Since $P$ is input-consistent and by hypothesis, 
$\ol v$ is linear and has {\em variables} in the free positions, 
and variables or flat terms in the controlled positions. Moreover, 
since $P$ is simple and $A$ is selectable, ${\ol s}$ is non-variable
in the controlled positions. Considering in addition that the clause
is a fresh copy renamed apart from $Q$, it follows that $\ol s$ is
an instance of $\ol v$. Let $\theta_1$ be the substitution
with $\Dom(\theta_1) \subseteq \Var({\ol v})$ 
such that $\ol v\theta_1 = \ol s$.

Since $\ol t$ is a linear vector of variables, there is
a substitution $\theta_2$ such that 
$\Dom(\theta_2) \subseteq \Var(\ol t)$ and
$\ol t \theta_2 = \ol u \theta_1$.

Since $Q$ is simply moded, we have 
$\Var({\ol t}) \cap \Var({\ol s}) = \emptyset$, and therefore
$\Var({\ol t}) \cap \Var({\ol v} \theta_1) = \emptyset$.
Thus it follows by the previous paragraph 
that $\theta = \theta_1 \theta_2$ is
an MGU of
$p({\ol s},{\ol t})$ and $p({\ol v},{\ol u})$. 
More precisely, we have
$\ol s \theta_1\theta_2 = \ol s$, 
$\ol v \theta_1\theta_2 = \ol v\theta_1$,
$\ol u \theta_1\theta_2 = \ol u\theta_1$, and
$\ol t \theta_1\theta_2 = \ol t\theta_2$, and so in particular, 
the derivation step using $\theta$ is input-consuming. 
Since mgu's are unique modulo renaming, the first statement follows.

We now show the second statement. If $H$ contains flat (i.e.,
non-variable) terms in all controlled positions, then clearly $A$ must
be non-variable in those positions for the derivation step using the
clause to be input-consuming. But then $A$ is also selectable.
Since the same holds for every clause, the statement follows.
\end{proof}
\end{rmlemma}

In order to assess how realistic these conditions are, we have checked
them against a number of programs from various collections. (The results
can be found in Sec.~\ref{app:benchmarks}).  Concerning the statement that
all delay-respecting derivations are input-consuming, we are convinced
that this is the case in the overwhelming majority of practical cases.
Concerning the converse, that is, that all input-consuming derivations
are delay-respecting, we could find different examples in which this
was not the case. In many of them this could be fixed by a simple
transformation of the programs\footnote{To give an intuitive idea, the
  transformation would, e.g., replace the clause
  \texttt{even(s(s(X))):- even(X).}\ with \texttt{even(s(Y)):-
    s\_decomp(Y,X), even(X).}, where we define
  \texttt{s\_decomp(s(X),X).} and the mode is
  \texttt{s\_decomp(In,Out).}},  in other cases it could not (e.g.,
\texttt{flatten}, \cite{SS86}).  Nevertheless, we strongly believe
that the latter form a small minority.

The delay declarations for the considered programs were either given
or derived based on the presumed mode. Note that delay declarations as
in Def.\ \ref{def:simple} can be more efficiently implemented than,
e.g., delay declarations testing for groundness. Usually, the
derivations permitted by the latter delay declarations are a strict
subset of the input-consuming derivations.

\section{Properties of Input-Consuming Derivations}\label{sec:IC-prop}
This section contains some technical results about input-consuming
derivations that are needed in the proofs. Moreover, it defines 
{\em simply-local substitutions}, which are crucial for our
semantics.

\subsection{Switching and Pushing}
The material in this subsection is not contained in the conference
version~\cite{BERS01}, and is needed in the proofs but not for
understanding the main results of this paper.

We recall the following results~\cite{BER99}.
Notice that they have been proven for nicely-moded programs and queries.
However,  since  simply-modedness is a special case of
nicely-modedness, the properties stated below also
apply to the class of programs and queries  considered in this paper.

The following lemma states that the only variables of a nicely-moded
query that can be ``affected'' through the computation of
an input-consuming derivation with  a nicely-moded
 program are those occurring in some output positions.

\begin{rmlemma} 
\label{lem:prop-in*}
Let the program $P$ and the query ${\ol A}$ be nicely-moded.  Let also
$ {\ol A}\stackrel{\theta}  \Dersym {\ol A}'$ be a partial
input-consuming derivation of $P\cup\{{\ol A}\}$.  Then, for all $x\in {\it
  Var}({\ol A})$ and $x\not \in{\it Var}({\it Out}({\ol A}))$, 
$x\theta=x$.
\end{rmlemma}

The following definition is due to Smaus \cite{Smaus-thesis}.

\begin{rmdefinition}
  Let $\mathbf{ A},B,\mathbf{ C} \stackrel{\theta}\Longrightarrow
  (\mathbf{ A},\mathbf{ B},\mathbf{ C})\theta$ be a derivation step.
  We say that each atom in $\ol B\theta$ is a \emph{direct descendant
    of B}, and for each atom $A$ in $(\ol A,\ol C)$, $A\theta$ is a
  \emph{direct descendant of A}. We say that $A$ is a descendant of
  $B$ if the pair $(A,B)$ is in the reflexive, transitive closure of
  the relation \emph{is a direct descendant}.
Consider a derivation 
$Q_0\stackrel{\theta_1}\Longrightarrow\cdots
\stackrel{\theta_i}\Longrightarrow Q_i \cdots
\stackrel{\theta_j}\Longrightarrow Q_j\stackrel{\theta_{j+1}}\Longrightarrow 
Q_{j+1}\cdots$. We say that $ Q_j\stackrel{\theta_{j+1}}\Longrightarrow 
Q_{j+1}\cdots$ is a $\ol B$-step if $\ol B$ is a subquery of $Q_i$ and
the selected atom in $Q_j$ is a descendant of an atom in $\ol B$.
\qed
\end{rmdefinition}

The next corollary is an immediate consequence of the
Left-Switching Lemma \cite{BER99}.

\begin{rmcoro}
\label{cor-sw}
Let the program $P$ and the query $\mathbf{ A},\mathbf{ B}$ be nicely-moded.
Suppose that
\[\delta: \mathbf{ A},\mathbf{ B}\stackrel{\theta}
\longmapsto \mathbf{ C}\]
is a partial input-consuming derivation of $P\U\{\ol A,\ol B\}$.
Then there exist $\ol C_1$ and $\ol C_2$ and a partial input-consuming derivation
\[\mathbf{ A},\mathbf{ B}\stackrel{\theta_1}
\longmapsto \mathbf{ C}_1, \mathbf{ B}\theta_1 \stackrel{\theta_2}
\longmapsto \mathbf{ C}_1,\mathbf{ C}_2\]
such that 
$\ol C=\ol C_1,\ol C_2$, $
\;\theta=\theta_1\theta_2$,
all the $\mathbf{ A}$-steps are performed in  the prefix 
$\mathbf{ A},\mathbf{ B}\stackrel{\theta_1}
\longmapsto~\mathbf{ C}_1, \mathbf{ B}\theta_1 $
and all the $\mathbf{ B}$-steps are performed in  the
suffix $ \mathbf{ C}_1, \mathbf{ B}\theta_1 \stackrel{\theta_2}
\longmapsto \mathbf{ C}_1,\mathbf{ C}_2$.
\qed
\end{rmcoro}

\begin{rmlemma}\textbf{(Input Pushing Lemma)}\ \label{input-pushing-lemma} 
Let the program $P$ and the query $\ol A$ be nicely-moded.
Let $\theta$ be a substitution such that 
$\mathit{Var}(\theta) \cap \mathit{Var}(\mathit{Out}({\ol A})) = \emptyset$.
Then
for every (partial) input-consuming 
derivation  $\delta: \ol A
\stackrel{\sigma}\longmapsto \ol B$,
there exists a (partial) input-consuming 
derivation  $\delta': \ol A \theta
\stackrel{\sigma'}\longmapsto \ol B'$
 such that 
\begin{itemize}
\item they  have the same length,
\item for
every derivation step, atoms in the same positions are selected and
the input clauses employed are variants of each other.
\end{itemize}
Moreover, $\ol A\theta\sigma'$ is an instance of $\ol A\sigma$
and $\ol B'$ is an instance of $\ol B$. 

\begin{proof} Notice that
since 
$\mathit{Var}(\theta) \cap \mathit{Var}(\mathit{Out}({\ol A})) =
\emptyset$  
by hypothesis, 
$\ol A\theta$ is nicely-moded as well. Since, by Lemma 
\ref{lem:prop-in*},
input-consuming derivations only affect variables
occurring in the output positions of a query,
one only has to appropriately instantiate every resolvent
in the derivation.
Clearly, every resolution step remains input-consuming
(the selected atom is just instantiated a bit further).
\end{proof}
\end{rmlemma}

\subsection{Simply-Local Substitutions}
We now define \emph{simply-local} substitutions,
which reflect the way clauses become instantiated in
input-consuming derivations. A simply-local substitution can be
decomposed into several substitutions, corresponding to the
instantiation of the \emph{output} of each \emph{body atom}, as well
as the \emph{input} of the \emph{head}.  

\begin{rmdefinition}
\label{def:simply-local}
Let $\theta$ be a substitution. We say that 
$\theta$ is \emph{simply-local} wrt.\ 
the clause 
$c: p({\ol t}_0,{\ol s}_{n+1})\la p_1({\ol s}_1,{\ol
    t}_1),\ldots,p_n({\ol s}_n,{\ol t}_n)$
 iff 
there exist substitutions $\sigma_0,\sigma_1\ldots,\sigma_n$
and disjoint sets of fresh (wrt.\  $c$) variables
$v_0,v_1,\ldots,v_n$   
such that
 $\theta=\sigma_0\sigma_1\cdots\sigma_n$ 
where for $i\in \{0,\ldots,n\}$,
\begin{itemize}
\item $\mathit{Dom}(\sigma_i)\subseteq \mathit{Var}({\ol t}_i)$,
\item  $\mathit{Ran}(\sigma_i)\subseteq  
\mathit{Var}({\ol s}_i \sigma_0\sigma_1\cdots\sigma_{i-1}) \cup
v_i$.
\end{itemize}

$\theta$ is \emph{simply-local} wrt.\ a query $\ol B$ iff
  $\theta$ is simply-local wrt.\ the clause $q \la \ol B$ where $q$ is
  any variable-free atom. \qed
\end{rmdefinition}

We make two remarks about this definition.

\begin{remark}\label{rem:simply-local}\ 
\begin{enumerate}
\item \label{abuse}
Concerning the case $i=0$, 
the term vector ${\ol s}_0$ does not exist, but by abuse of notation, 
we postulate
$\mathit{Var}({\ol s}_0 \dots) = \emptyset$.
\item \label{sigma0-remark}
In the case of a simply-local substitution wrt.\ a query, $\sigma_0$ is
the empty substitution, since 
$\mathit{Dom}(\sigma_0)\subseteq \mathit{Var}(q)$ where $q$ is an (imaginary)
variable-free atom.
\end{enumerate}
\end{remark}

\begin{myexample}
Consider \texttt{APPEND} in mode \texttt{append(In,In,Out)}, and its
recursive clause 
$c: \tt append([H|Xs],Ys,[H|Zs]) \la append(Xs,Ys,Zs)$.
The substitution 
$\theta = \{\tt H/V, Xs/[],$ $\tt Ys/[W], Zs/[W]\}$ is simply-local wrt.\ $c$: let
$\sigma_0 = \{\tt H/V, Xs/[], Ys/[W]\}$ and
$\sigma_1 = \{\tt Zs/[W]\}$; then
$\mathit{Dom}(\sigma_0)\subseteq \{\tt H, Xs, Ys\}$, and
$\mathit{Ran}(\sigma_0)\subseteq v_0$ where 
$v_0 = \{\tt V, W\}$, and
$\mathit{Dom}(\sigma_1)\subseteq \{\tt Zs\}$, and
$\mathit{Ran}(\sigma_1)\subseteq 
  \mathit{Var}({\tt (Xs,Ys)}\sigma_0)$.
\end{myexample}

From the next lemma, we will be able to conclude that if 
${\ol A},B,{\ol C}
\stackrel{\theta}\Longrightarrow
({\ol A},{\ol B},{\bf C})\theta$ 
is an input-consuming derivation step 
using clause
$c: H\leftarrow {\ol B}$, then 
without loss of generality, $\theta$ can be decomposed into two
substitutions that are simply-local wrt.~the clause $H\la$ and 
the query $B$, respectively.

\begin{rmlemma}\textbf{(Simply-Local MGU)}\ 
\label{lemma:simply-local-mgu}
Let the atoms $A$ and $H$ be variable disjoint and  $A$ be
simply-moded.
Suppose  that there exists $\vartheta={\it mgu}(A,H)$ such that
${\it In}(A\vartheta)= {\it In}(A)$.
Then there exist two substitutions $\sigma^H_0$ and $\sigma^A_1$ 
such that $\sigma^H_0 \sigma^A_1 ={\it mgu}(A,H)$, and 
$\sigma^H_0$ is simply-local wrt.\ the clause $H\leftarrow$, and 
$\sigma^A_1$ is simply-local wrt.\ the query $A$.

\begin{proof}
Let $A=p(\ol s,\ol t)$ and $H=p(\ol v,\ol u)$.
By properties of mgu's (see \cite[Corollary 2.25]{Apt97}), 
there exist substitutions $\sigma^H_0$ and $\sigma^A_1$
(the names have been chosen to correspond closely to
Def.~\ref{def:simply-local}) such that 
\[
\sigma^H_0=\mathit{mgu}(\ol s, \ol v), \quad 
\sigma^A_1=\mathit{mgu}(\ol t\sigma^H_0, \ol u\sigma^H_0)\quad \mbox{and} \quad
\sigma^H_0 \sigma^A_1 ={\it mgu}(A,H),
\]
and all those mgu's are relevant.
Since, by hypothesis, $\ol v \vartheta =\ol s\vartheta=\ol s$,
it follows that $\ol s$ is an instance of $\ol v$. Hence, we can
assume without loss of generality that 
$\sigma^H_0$ is such that $\ol v\sigma^H_0=\ol s$ and thus
\begin{itemize}
\item
$\mathit{Dom}(\sigma^H_0)\subseteq \mathit{Var}(\ol v)$, and
\item 
$\mathit{Ran}(\sigma^H_0)\subseteq \mathit{Var}(\ol s)$.
\end{itemize}
Since $\mathit{Var}(\ol s)$ is fresh wrt.~$H$, this means that 
$\sigma^H_0$ is simply-local wrt.\ the clause $H\leftarrow$.

By relevance of $\sigma^H_0$, simply-modedness of $A$ and the fact
that $A$ and $H$ are variable disjoint, it follows that
$\mathit{Dom}(\sigma^H_0)\cap \mathit{Var}(\ol t)=\emptyset$.  Hence,
$\sigma^A_1=\mathit{mgu}(\ol t, \ol u\sigma^H_0)$.  Since, by
simply-modedness of $A$, $\ol t$ is sequence of distinct variables, we
can assume without loss of generality that 
$\sigma^A_1$ is such that $\ol t\sigma^A_1=\ol u\sigma^H_0$ and
thus
\begin{itemize}
\item $\mathit{Dom}(\sigma^A_1)\subseteq 
\mathit{Var}(\ol t)$, and
\item
$\mathit{Ran}(\sigma^A_1)\subseteq 
\mathit{Var}(\ol u\sigma^H_0)
\subseteq
 \mathit{Ran}(\sigma^H_0) \cup \mathit{Var}(\ol u)\subseteq
\mathit{Var}(\ol s) \cup \mathit{Var}(\ol u)$.
\end{itemize}
Since $\mathit{Var}(\ol u)$ is fresh wrt.~$A$ and noting
Remark~\ref{rem:simply-local}~(\ref{sigma0-remark}), 
this means that 
$\sigma^A_1$ is simply-local wrt.\ the query $A$.
\end{proof}
\end{rmlemma}

\begin{remark}\label{rem:rel-mgu-sub}
From now on we assume that all the mgu's used in input-consuming
derivations are composed of two simply-local substitutions as in
Lemma~\ref{lemma:simply-local-mgu}.
\end{remark}

The following lemma shows how the accumulated substitution of a
derivation starting in a query $A_1,\ldots,A_n$
can be decomposed into $n$ substitutions each 
corresponding to one atom $A_i$.

\begin{rmlemma}
\label{lemma:ic-cas}
Let the program $P$ and the query $A_1,\ldots,A_n$ be simply-moded.
Suppose that $\delta: A_1,\ldots,A_n\stackrel{\vartheta}  \Dersym \ol A$ is an
input-consuming partial derivation of $P\cup\{A_1,\ldots,A_n\}$. Then,
 there exist $\sigma_1,\ldots,\sigma_n$ substitutions
and $v_1,\ldots,v_n$  disjoint sets of fresh variables
(wrt.\ $A_1,\ldots,A_n$)
such that
$\vartheta=\sigma_1\cdots\sigma_n$  and
\begin{itemize}
\item  for $i\in \{1,\ldots,n\}$,
 $\mathit{Dom}(\sigma_i)\subseteq \mathit{Var}(\mathit{Out}(A_i))\cup v_i$,
\item   for $i\in \{1,\ldots,n\}$,
$\mathit{Ran}(\sigma_i)\subseteq \mathit{Var}(\mathit{In}(A_i\sigma_1\cdots\sigma_{i-1})) \cup v_i$.
\end{itemize}

\begin{proof}
By induction on $n$.

\emph{Base}.  Let $n=1$. In this case,
$
\delta: A\stackrel{\vartheta}  \Dersym \ol A.
$
By Lemma \ref{lem:prop-in*}, there exists a set 
of variables $v$ such that $\mathit{Var}(c)\subseteq v$
 and $v \cap \mathit{Var}(A)=\emptyset$ (i.e.\ $v$ is 
fresh w.r.t.\ $A$), and
$\mathit{Dom}(\vartheta)\subseteq \mathit{Var}(\mathit{Out}(A))\cup v$.

Suppose that $\delta$ is of the form
$$A\stackrel{\vartheta_1}  \Longrightarrow (B_1,\ldots,B_m)\vartheta_1
\stackrel{\vartheta_2} \Dersym \ol A$$
where $c: H\leftarrow B_1,\ldots,B_m$ is the input clause used 
in the first derivation step, $\vartheta_1={\it mgu}(A,H)$
such that ${\it In}(A\vartheta_1)= {\it In}(A)$ and $\vartheta=
\vartheta_1\vartheta_2$.
Since $\vartheta_1$ is simply-local wrt.\ $A$ and $H$,
we have that $\vartheta_1=\sigma_1\sigma_2$ where
\begin{itemize}
\item
$\mathit{Dom}(\sigma_1)\subseteq \mathit{Var}(\mathit{In}(H))$,
\item $\mathit{Ran}(\sigma_1)\subseteq \mathit{Var}(\mathit{In}(A))$,
\item $\mathit{Dom}(\sigma_2)\subseteq
 \mathit{Var}(\mathit{Out}(A))$,
\item $\mathit{Ran}(\sigma_2)\subseteq
\mathit{Var}(\mathit{In}(A)) \cup  \mathit{Var}(\mathit{Out}(H))$.
\end{itemize}

Now 
$\mathit{Var}(\vartheta_1)\subseteq
 \mathit{Var}(\mathit{In}(A))\cup v$.
By standardisation apart and simply-modedness of $A$, it follows that
$H\sigma_1\sigma_2=H\sigma_1$, and so $(B_1,\ldots,B_m)\vartheta_1=
(B_1,\ldots,B_m)\sigma_1$. 
Consider  the derivation $ (B_1,\ldots,B_m)\sigma_1
\stackrel{\vartheta_2} \Dersym~\ol A$. 
We have that
$\mathit{Var}(\vartheta_2)\subseteq \mathit{Var}(\sigma_1)\cup v$.
Since $\mathit{Var}(\sigma_1)\subseteq \mathit{Var}(\mathit{In}(A))
\cup \mathit{In}(H))$, we have that
$\mathit{Var}(\vartheta_2)\subseteq
 \mathit{Var}(\mathit{In}(A))\cup v$.

Thus, $\mathit{Ran}(\vartheta)=\mathit{Ran}(\vartheta_1\vartheta_2)
\subseteq \mathit{Var}(\vartheta_1)\cup
\mathit{Var}(\vartheta_2)\subseteq \mathit{Var}(\mathit{In}(A))\cup v$.

\emph{Induction step}.  Let $n>1$.
By Corollary \ref{cor-sw}, there exist
$\sigma_1,\ldots,\sigma_n$
 substitutions
 such that
 $$
\begin{array}{ll}
\delta: A_1,\ldots,A_n\stackrel{\sigma_1}\Dersym 
\ol C_1,(A_2,\ldots,A_n)\sigma_1  &  
\stackrel{\sigma_2}\Dersym \cdots
 (\ol C_1,\ldots,\ol C_{n-1}),A_n\sigma_1\cdots \sigma_{n-1}\\
                                  &   
\stackrel{\sigma_n}  \Dersym \ol C_1,\ldots ,\ol C_{n}
\end{array}
$$
such that $\ol A=\ol C_1,\ldots, \ol C_n$, and
$\vartheta=\sigma_1\cdots,\sigma_n$, and all the $A_i$-steps are performed in the
sub-derivation 
$$
\ol C_1,\ldots, \ol C_{i-1},(A_i,\ldots,A_n)\sigma_1\cdots
\sigma_{i-1}\stackrel{\sigma_i}  \Dersym
 \ol C_1,\ldots, \ol C_{i},(A_{i+1},\ldots,A_n)\sigma_1\cdots
\sigma_{i}.
$$

By the induction hypothesis and standardisation apart,
there exist $v_1,\ldots,v_n$  disjoint sets of fresh variables
(wrt.\ $A_1,\ldots,A_n$)
such that
 for all $i\in \{1,\ldots,n\}$,
\begin{itemize}
\item 
 $\mathit{Dom}(\sigma_i)\subseteq \mathit{Var}(\mathit{Out}(A_i))\cup v_i$,
\item 
$\mathit{Ran}(\sigma_i)\subseteq \mathit{Var}(\mathit{In}(A_i\sigma_1\cdots\sigma_{i-1})) \cup v_i$.
\end{itemize}
\vskip -6mm

\end{proof}
\end{rmlemma}

\section{A Denotational Semantics}
\label{sec:semantics}
Previous declarative
semantics for logic programs cannot correctly model 
dynamic scheduling. E.g., none of them reflects the
fact that \texttt{append(X,Y,Z)} deadlocks.  
We define a model-theoretic semantics that models computed answer
substitutions of input-consuming derivations of simply-moded programs
and queries.

In predicate logic, an interpretation states which formulas are true
and which ones are not. For our purposes, it is convenient to
formalise this by defining an interpretation $I$ as a set of atoms
closed under variance. Based on this notion and simply-local
substitutions, we now define a restricted notion of model.

\begin{rmdefinition}
Let $M$ be an interpretation.
We say that $M$ is a \emph{simply-local model} of 
$c: H\leftarrow B_1,\ldots,B_n$ iff for every substitution 
$\theta$ simply-local wrt.\ $c$, 
\begin{equation}
\mbox{if $B_1\theta,\ldots,B_n\theta\in M$ then $ H\theta\in M$.}
\label{SL-model-eq}
\end{equation}
$M$ is a \emph{simply-local model} of a program $P$ iff it is a
  simply-local model of each clause of it.  \qed
\end{rmdefinition}

Note that a simply-local model 
is not necessarily a model in the classical sense, since 
the substitution in (\ref{SL-model-eq}) is required to be 
simply-local.
For example, given the program
$\{\tt q(1).,\; p(X) \!\la\! q(X).\}$
with modes 
$\tt q(In),\ p(Out)$, a model must contain the atom
$\tt p(1)$, whereas a simply-local model does not necessarily contain
$\tt p(1)$, since $\{\tt X/1\}$ is not simply-local wrt.\ 
$\tt p(X) \la q(X).$

We now show that there exists a minimal simply-local model and that it
is bottom-up computable. For this we need the following operator
$\mathit{T^{SL}_P}$ on interpretations: Given a program $P$
and an interpretation $I$, define
\[
\begin{array}{ll}
\mathit{T^{SL}_P}(I) = \{ H\theta \mid 
 & 
\exists\ c: H \la  B_1,\ldots,B_n \in P,\\
 & 
\exists\  \theta \ \mbox{simply-local wrt.}\ c,\\
 & B_1,\ldots,B_n\theta \in I\}.
\end{array}
\]
Operator's powers are defined in the standard way:
$\mathit{T^{SL}_P}\uparrow 0(I)  =  I$, 
$\mathit{T^{SL}_P}\uparrow (i+1)(I)  =  \mathit{T^{SL}_P}(\mathit{T^{SL}_P}
\uparrow i(I))$, and 
$\mathit{T^{SL}_P}\uparrow \omega (I)  =  \bigcup_{i=0}^\infty \mathit{T^{SL}_P}
\uparrow i(I)$.
It is easy to show that $\mathit{T^{SL}_P}$ is continuous on the
lattice where interpretations are ordered by set inclusion. Hence, by
well-known results~\cite{Apt97}, 
$\mathit{T^{SL}_P}\uparrow \omega$ exists and is
the least fixpoint of $\mathit{T^{SL}_P}$. 

\subsection{Modelling Complete Derivations}
\label{subsec:complete}
In this subsection, we use least simply-local models for describing
the usual complete derivations. 
As suggested above, $\mathit{T^{SL}_P}$ can be used to compute the
least simply-local model of a program.

\begin{propo}\label{propo:least-model-exists}
  Let  $P$ be simply-moded. Then 
$\mathit{T^{SL}_P}\uparrow \omega (\emptyset)$ is the least
simply-local model of $P$. 
\qed\end{propo}

We denote the least
simply-local model of $P$ by $\mathit{M^{SL}_P}$.

The following lemma is a special case of the statement that our 
semantics is correct, fully abstract and compositional. 
It is needed in the proof of the subsequent theorem, and is not
included in~\cite{BERS01}.

\begin{rmlemma}
\label{lem:atomic-sem-total}
Let the program 
$P$ and the atom $A$ be simply-moded.
The following statements are equivalent:
\begin{itemize}
\item[(i)]  there exists an input-consuming successful derivation
$A\stackrel{\vartheta}  \Dersym_P~\Box$,
\item[(ii)]  there exists 
a substitution $\theta$ such that $A\theta\in \mathit{M^{SL}_P}$ and 
$\mathit{In}(A\theta)=\mathit{In}(A)$,
\end{itemize}
where $A\vartheta$ and $A\theta$ are variant.

\begin{proof}

$\mathrm{(i)}\Rightarrow \mathrm{(ii)}.$
By induction on the length of $\delta$.


\emph{Base}.  Let $\mathit{len}(\delta)=1$.
In this case $\delta$   has  the form
$$A\stackrel{\vartheta}  \Longrightarrow_P \Box$$
where
$c: H\leftarrow$ is  the input clause and
$\vartheta={\it mgu}(A,H)$ satisfies ${\it In}(A\vartheta)={\it In}(A)$.
Since $\vartheta$ is simply-local wrt.\ $A$ and $H$,
by Remark \ref{rem:rel-mgu-sub},
 $\vartheta_{|H}$ is simply-local wrt.\ $ H\leftarrow$.
 Hence, by definition of
$\mathit{T^{SL}_P}$, 
\[
H\vartheta_{|H}=H\vartheta=A\vartheta\in \mathit{T^{SL}_P}\uparrow 1
(\emptyset)
\subseteq \mathit{M^{SL}_P}.
\]

\emph{Induction step}.   Let $\mathit{len}(\delta)>1$.
In this case, $\delta$ has the form 
$$A\stackrel{\vartheta_1}  \Longrightarrow (B_1,\ldots,B_n)\vartheta_1
\stackrel{\vartheta_2} \Dersym \Box$$
where $c: H\leftarrow B_1,\ldots,B_n$ is the input clause used 
in the first derivation step, $\vartheta_1={\it mgu}(A,H)$
satisfies ${\it In}(A\vartheta_1)= {\it In}(A)$ and $\vartheta=
\vartheta_1\vartheta_2$.
Since $\vartheta_1$ is simply-local wrt.\ $A$ and $H$,
by Remark \ref{rem:rel-mgu-sub},
 $\vartheta_{1|H}$ is simply-local wrt.\ $ H\leftarrow$.
Let $\vartheta_{1|H}=\sigma_0$.
 By Definition \ref{def:simply-local} of
simply-local substitution, there exists a set $v_0$ of fresh variables
(wrt.\ $c$) such that
\begin{equation}
\label{eq:doms0}
\mathit{Dom}(\sigma_0)\subseteq \mathit{Var}(\mathit{In}(H))
\end{equation}
and
\begin{equation}
\label{eq:rans0}
\mathit{Ran}(\sigma_0)\subseteq v_0.
\end{equation}
By standardisation apart,
\begin{equation}
\label{eq:nos'0}
(B_1,\ldots, B_n)\vartheta_1=(B_1,\ldots, B_n)\sigma_0.
\end{equation}
By (\ref{eq:nos'0}) and the Left-Switching Lemma, there exist
$\sigma'_1,\ldots,\sigma'_n$ and a derivation
$\delta'$ isomorphic to $\delta$ (modulo the Left-Switching Lemma),
and $\delta'$ has the form 
$$A\stackrel{\vartheta_1}  \Longrightarrow (B_1,\ldots,B_n)\sigma_0
\stackrel{\sigma'_1}  \Dersym (B_2,\ldots,B_n)\sigma_0\sigma'_1\cdots
B_n\sigma_0\sigma'_1\cdots\sigma'_{n-1}
\stackrel{\sigma'_n} \Dersym \Box$$
where  $\vartheta_2=\sigma'_1\cdots\sigma'_n$.
  In particular, for all $i\in\{1,\ldots,n\}$,
$\delta_i: B_i\sigma_0\sigma'_1\cdots\sigma'_{i-1}\stackrel{\sigma'_i} \Dersym \Box$ 
is  an input-consuming successful derivation which is strictly shorter than $\delta$.

Hence, by the inductive hypothesis,
for all $i\in\{1,\ldots,n\}$,
\begin{equation}
\label{eq:ind1}
 B_i\sigma_0\sigma'_1\cdots\sigma'_i\in \mathit{M^{SL}_P}.
\end{equation}

By simply-modedness of $c$,
${\it Out}((B_1,\ldots ,B_n)\sigma_0)={\it Out}(B_1,\ldots ,B_n)$.

By Lemma \ref{lemma:ic-cas},
 there exist distinct sets of fresh variables $v_1,\ldots,v_n$,  
such that \linebreak
${\it Dom}(\sigma'_i)\subseteq 
{\it Var}({\it Out}(B_i\sigma'_1\cdots \sigma'_{i-1}))\cup v_i$.
By induction on $i$, one can prove that,  for all $i\in\{1,\ldots,n\}$, 
\begin{equation}
\label{eq:dom-si-ref}
{\it Dom}(\sigma'_i)\subseteq {\it Var}({\it Out}(B_i))\cup v_i.
\end{equation}

The base case is trivial.
The induction step, follows 
 from the inductive hypothesis
(i.e., ${\it Dom}(\sigma'_j)\subseteq {\it Var}({\it Out}(B_j))\cup v_j$
 for $j\in\{1,\ldots,i-1\}$), standardisation apart
 and
simply-modedness of~$c$.

For all  $i\in \{1,\ldots,n\}$, let $\sigma_i=\sigma'_{i|\mathit{Var}(\mathit{Out}(B_i))}$. 
Hence, by (\ref{eq:dom-si-ref}),
\begin{equation}
\label{eq:dom-si}
{\it Dom}(\sigma_i)\subseteq {\it Var}({\it Out}(B_i)).
\end{equation}

By standardisation apart,
for all $i\in\{1,\ldots,n\}$, 
\begin{equation}
\label{eq:body3}
B_i\sigma_0\sigma'_1\cdots\sigma'_i=
B_i\sigma_0\sigma_1\cdots\sigma_i.
\end{equation} 

By Lemma \ref{lemma:ic-cas} and (\ref{eq:body3}),
for  all $i\in \{1,\ldots,n\}$,
\begin{equation}
\label{eq:ran-si}
\mathit{Ran}(\sigma_i)\subseteq \mathit{Var}(\mathit{In}
(B_i\sigma_0\sigma_1\cdots \sigma_{i-1}))\cup v_i.
\end{equation}

By standardisation apart and simply-modedness of $c$,
 it follows that for all
 $j\in\{i+1,\ldots,n\}$,
 ${\it Var}(B_i\sigma_0\sigma_1\cdots \sigma_i)\cap
{\it Dom}(\sigma'_j)=\emptyset$. Hence, by (\ref{eq:body3}),
it  follows that
\begin{equation}
\label{eq:body4}
B_i\sigma_0\sigma'_1\cdots\sigma'_i=B_i\sigma_0\sigma_1\cdots\sigma_n.
\end{equation}

By (\ref{eq:doms0}), (\ref{eq:rans0}), (\ref{eq:dom-si}), (\ref{eq:ran-si})
 and the fact that $v_0,v_1,\ldots,v_n$ are disjoint sets of variables,
it follows that

\begin{equation}
\label{eq:simply-localsub}
\sigma_0\sigma_1\cdots\sigma_n \mbox{ is simply local wrt.\ } c.
\end{equation}

Moreover, by (\ref{eq:ind1}) and (\ref{eq:body4}),
\begin{equation}
\label{eq:body}
B_i\sigma_0\sigma_1\cdots\sigma_n\in \mathit{M^{SL}_P}.
\end{equation}

By  definition of
$\mathit{T^{SL}_P}$,  $H\sigma_0\sigma_1\cdots\sigma_n \in \mathit{M^{SL}_P}$.
Since $A\vartheta=A\vartheta_1\vartheta_2=H\vartheta_1\vartheta_2=
H\sigma_0\vartheta_2=H\sigma_0\sigma'_1\cdots\sigma'_n=H\sigma_0\sigma_1\cdots\sigma_n$, 
we have proven that
\begin{equation}
\label{eq:fin}
 A\vartheta \in \mathit{M^{SL}_P}.
\end{equation}
Since by Lemma \ref{lem:prop-in*}, $\mathit{In}(A\vartheta)=\mathit{In}(A)$,
this completes the proof of the 
``$\mathrm{(i)}\Rightarrow \mathrm{(ii)}$'' direction.

$\mathrm{(ii)}\Rightarrow \mathrm{(i)}.$
We first need to establish the following fact.

\begin{fact} 
\label{mgu-ic}
Let the atom $A$ and the clause $c: H\leftarrow B_1,\ldots,B_n$
be simply-moded.  Suppose that there exist
two substitutions $\sigma$ and $\theta$ such that
\begin{itemize}
\item  $\sigma$ is simply-local wrt.\ $c$,
\item $A\theta= H\sigma$,
\item  $\mathit{In}(A)=\mathit{In}(A\theta)$.
\end{itemize}
Then, 
for each variant $c': H'\leftarrow B'_1,\ldots,B'_n$  of $c$ variable 
disjoint with  $A$, 
there exists $\vartheta={\it mgu}(A,H')$ such that
 $A\vartheta= A\theta$ and 
$\mathit{In}(A)= \mathit{In}(A\vartheta)$.
\end{fact}
\begin{prooffact} 
Since
 $A\theta= H\sigma$, it follows that
(since $A$ and $H'$ are variable-disjoint) 
$A$ and $H'$ are unifiable, and
$A\theta $ ($= H\sigma$) is an instance of the most general common
instance of $A$ and $H'$. Now, since by assumption
$\mathit{In}(A)=\mathit{In}(A\theta)$ and
$\sigma$ is simply-local wrt.\ $c$, we can choose 
$\vartheta={\it mgu}(A,H')$ such that
\begin{eqnarray}
\label{eq:ina}
\mathit{In}(A) &=& \mathit{In}(A\vartheta).\\
\label{eq:outH}
\mathit{Out}(H'\vartheta)&=& \mathit{Out}(H\sigma).
\end{eqnarray}
Using that $\vartheta$ is an mgu,
the assumptions in the statement, 
and (\ref{eq:ina}), we have
 ${\it In}(H'\vartheta)= {\it In}(A\vartheta)=
{\it In}(A)={\it In}(A\theta)= {\it In}(H\sigma)$, i.e.,
\begin{equation}
\label{eq:inH}
\mathit{In}(H'\vartheta)= \mathit{In}(H\sigma).
\end{equation}
By (\ref{eq:outH}) and (\ref{eq:inH}),
\begin{equation}
\label{eq:eqA}
  A\vartheta=H'\vartheta= H\sigma= A\theta.
\end{equation}
By (\ref{eq:ina}) and (\ref{eq:eqA}),
$\mathit{In}(A)=\mathit{In}(A\theta)$.
This completes the proof of the Fact.
\end{prooffact}

We now continue the proof of the main statement.
We show by induction on $i$ 
that if $A\theta\in \mathit{T^{SL}_P}\uparrow i
(\emptyset)$
for some $i>0$ and
substitution $\theta$ such that $\mathit{In}(A)=\mathit{In}(A\theta)$,
then there exists 
 an input-consuming successful derivation 
$$
A\stackrel{\vartheta} \Dersym_P \Box
$$ 
and $A\vartheta=A\theta$.

{\it Base}. Let $i=1$. In this case,
 $A\theta\in \mathit{T^{SL}_P}\uparrow 1 (\emptyset)$. 
By Definition  of $\mathit{T^{SL}_P}$,
there exists a clause  $c: H\leftarrow $  of $P$ and 
a substitution $\sigma$ such that $\sigma$ is
simply-local wrt.\ $c$ and $A\theta=H\sigma$.
Let  $H'\leftarrow $ be a variant of $c$ variable disjoint from
$A$. 
By Fact \ref{mgu-ic},
there exists an mgu $\vartheta$ of $A$ and $H$ such that $A\theta=
A\vartheta$ and $\mathit{In}(A)=\mathit{In}(A\vartheta)$, 
i.e., there exists an input-consuming
successful derivation
$A\stackrel{\vartheta} \Dersym_P
\Box$. 

\emph{Induction step}.
Let $i>1$ and $A\theta\in\mathit{T^{SL}_P}\uparrow i (\emptyset)$.
By definition of $\mathit{T^{SL}_P}$,
 there exists a clause  $c: H\leftarrow B_1,\ldots ,B_n$ 
 of  $P$ and
a substitution $\sigma$ such that
$\sigma$ is simply-local wrt.\ $c$,
 $(B_1,\ldots,B_n)\sigma\in 
\mathit{T^{SL}_P}\uparrow (i-1)$
and $A\theta= H\sigma$.

By Definition \ref{def:simply-local} of simply-local substitution,
there exist $\sigma_0,\sigma_1,\ldots,\sigma_n$ substitutions
and $v_0,v_1,\ldots,v_n$  disjoint sets of fresh  variables (wrt.\ $c$)
such that
 $\sigma=\sigma_0\sigma_1\cdots\sigma_n$ 
where
\begin{itemize}
\item $\mathit{Dom}(\sigma_0)\subseteq \mathit{Var}(\mathit{In}(H))$,
\item  $\mathit{Ran}(\sigma_0)\subseteq  v_0$,
\item  for $j\in \{1,\ldots,n\}$,
 $\mathit{Dom}(\sigma_j)\subseteq \mathit{Var}(\mathit{Out}(B_j))$,
\item   for $j\in \{1,\ldots,n\}$,
$\mathit{Ran}(\sigma_j)\subseteq 
\mathit{Var}(\mathit{In}(B_j \sigma_0\sigma_1\cdots\sigma_{j-1})) \cup v_j$.
\end{itemize}
By simply-modedness of $c$, we have that
 for all $j\in\{1,\ldots,n\}$,
$B_j\sigma_0\sigma_1\cdots\sigma_j=
B_j\sigma_0\sigma_1\cdots\sigma_n$.

So we have 
$(B_j\sigma_0\sigma_1\cdots\sigma_{j-1})\sigma_j\in
\mathit{T^{SL}_P}\uparrow (i-1)$
and
\[
\mathit{In}(B_j\sigma_0\sigma_1\cdots\sigma_{j-1}) = 
\mathit{In}((B_j\sigma_0\sigma_1\cdots\sigma_{j-1})\sigma_j),
\]
and hence, by the inductive hypothesis, for all $j\in\{1,\ldots,n\}$,
 there is an
input-consuming 
derivation 
$
B_j\sigma_0\sigma_1\cdots\sigma_{j-1}
\stackrel{\sigma'_j} \Dersym_P \Box
$
such that
$$
B_j\sigma_0\sigma_1\cdots\sigma_{j-1}\sigma'_j= B_j
\sigma_0\sigma_1\cdots\sigma_{j-1}\sigma_j.
$$

Let $ c': H'\leftarrow B'_1,\ldots ,B'_n$ be a variant of $c$
variable disjoint from $A$ such that $c'=c\rho$ for some renaming $\rho$.
By Fact \ref{mgu-ic}, there exists an mgu $\vartheta_1 $ of $A$ and $H'$ such that
 $\mathit{In}(A)= \mathit{In}(A\vartheta_1)$ and
$H'\vartheta_1=H\rho\vartheta_1=H\sigma_0$. 
Thus,
 $( B'_1,\ldots,B'_n)\vartheta_1=
(B_1,\ldots,B_n)\sigma_0.$ 
By the inductive hypothesis, for all $j\in{1,\ldots,n}$,
 there exists an
input-consuming successful derivation 
$B_j\sigma_0\sigma_1\cdots\sigma_{j-1}
\stackrel{\sigma'_j} \Dersym_P \Box$ such that
$B_j\sigma_0\sigma_1\cdots\sigma_{j-1}\sigma'_j= B_j
\sigma_0\sigma_1\cdots\sigma_{j-1}\sigma_j$.

Hence,
there exists an input-consuming successful derivation
$$(B_1,\ldots,B_n)\sigma_0 \stackrel{\vartheta_2} \Dersym_P~\Box$$
such that $(B_1,\ldots,B_n)\sigma_0\vartheta_2=
(B_1,\ldots,B_n)\sigma_0\sigma_1\cdots\sigma_n$.

 
Hence,
there exists an input-consuming  successful derivation
$$A \stackrel{\vartheta_1} \Longrightarrow_P
(B_1,\ldots,B_n)\vartheta_1 \stackrel{\vartheta_2} \Dersym_P~\Box$$
with $\vartheta=\vartheta_1\vartheta_2$
and 
$A\vartheta_1\vartheta_2=H\sigma_0\vartheta_2=H\sigma_0\sigma_1\cdots\sigma_n=H\sigma=A\theta$, 
i.e., $A\vartheta=A\theta$.
\end{proof}
\end{rmlemma}

We now state that our 
semantics is correct, fully abstract and compositional for complete derivations. 

\begin{rmtheorem}\label{thm:comp-sem-total}
Let the program $P$ and the query $\ol A$ be simply-moded.  The
following statements are equivalent:
\begin{itemize}
\item[(i)] there exists an input-consuming successful derivation $\ol
  A\stackrel{\vartheta} \Dersym_P~\Box$,
\item[(ii)] there exists a substitution $\theta$,
 simply-local wrt.~$\ol
  A$,
  such that 
$\ol A\theta\in \mathit{M^{SL}_P}$,
\end{itemize}
where $\ol A\theta$ is a variant of $\ol A\vartheta$.

\begin{proof}
Let $\ol A:=A_1,\ldots , A_n$.
The proof is by induction on  $n$.

\textit{Base}. Let $n=1$. In this case 
the thesis follows from Lemma \ref{lem:atomic-sem-total}.

\textit{Induction step}. Let $n>1$. 

$\mathrm{(i)}\Rightarrow \mathrm{(ii)}.$
Let $\theta=\vartheta_{|\ol A}$.
By Lemma \ref{lemma:ic-cas}, $\theta$ is simply-local wrt.\ $\ol A$.
By Corollary \ref{cor-sw}, there exists a successful input-consuming
derivation of the form
$$A_1,\ldots,A_n\stackrel{\vartheta_1}  \Dersym_P 
(A_2,\ldots,A_n)\vartheta_1\stackrel{\vartheta_2}  \Dersym_P\Box$$
where $\vartheta=\vartheta_1\vartheta_2$.

By Lemma \ref{lemma:ic-cas} and standardisation apart,
 $\mathit{Dom}(\vartheta_1)\subseteq
\mathit{Var}(\mathit{Out}(A_1))\cup v_1$ 
where $v_1$ is a set of fresh variables (wrt.\ $A_1,\ldots,A_n$),
and
$
\mathit{Dom}(\vartheta_2)\subseteq$ 
$\mathit{Var}(\mathit{Out}((A_2,\ldots,A_n)
\vartheta_1))\cup v, 
$ 
where $v$ is a set of fresh variables
(wrt.\ $A_1,\ldots,A_n$ and $\vartheta_1$).
By simply-modedness of $c$ and standardisation apart, 
 $\mathit{Var}(\mathit{Out}((A_2,\ldots,A_n)
\vartheta_1))=
\mathit{Var}(\mathit{Out}(A_2,\ldots,A_n))$, and so
$\mathit{Dom}(\vartheta_2)\subseteq$ 
$\mathit{Var}(\mathit{Out}(A_2,\ldots,A_n)
)\cup v$. 

By simply-modedness of $c$, $\mathit{Var}(A_1)\cap \mathit{Var}(\mathit{Out}(A_2,\ldots,A_n))=\emptyset$.
Hence, by standardisation apart,
$\mathit{Var}(A_1\vartheta_1)\cap \mathit{Dom}(\vartheta_2)=\emptyset$, i.e.,
 $A_1\vartheta_1\vartheta_2=A_1\vartheta_1$.

By the inductive hypothesis, $A_1\vartheta_1=A_1\vartheta_1\vartheta_2\in \mathit{M^{SL}_P}$ and
$(A_2,\ldots,A_n)\vartheta_1\vartheta_2\in
\mathit{M^{SL}_P}$, i.e., $\ol A\theta\in \mathit{M^{SL}_P}$.

$\mathrm{(ii)}\Rightarrow \mathrm{(i)}.$
By the inductive hypothesis, there exists an input-consuming successful derivation
$A_1\stackrel{\vartheta_1}  \Dersym_P \Box$ where
 $A_1\vartheta_1=A_1\theta$. Again
by the inductive hypothesis, there exists an input-consuming successful derivation
$(A_2,\ldots,A_n)\theta_{|A_1}\stackrel{\vartheta_2}  \Dersym_P \Box$\linebreak
such that 
$(A_2,\ldots,A_n)\theta_{|A_1}\vartheta_2=(A_2,\ldots,A_n)\,\theta$.
Since, by standardisation apart,\linebreak
$(A_2,\ldots,A_n)\theta_{|A_1}=(A_2,\ldots,A_n)\vartheta_1$, it follows that
there is an input-consuming successful derivation
$$A_2,\ldots,A_n\stackrel{\vartheta_1}  \Dersym_P 
(A_2,\ldots,A_n)\vartheta_1\stackrel{\vartheta_2}  \Dersym_P\Box.$$
\end{proof}
\end{rmtheorem}

\begin{myexample}\label{complete-model-ex}
Considering again \texttt{APPEND}, we have that
\[
\mathit{M^{SL}_{\tt APPEND}} = \bigcup_{n=0}^{\infty} \{ {\tt
  append}([t_1,\ldots,t_n],s,[t_1,\ldots,t_n|s])\ |\ t_1,\ldots,t_n,s
\mbox{ are any terms } \}.
\]
Using Theorem \ref{thm:comp-sem-total}, we can conclude that
the query \texttt{append([a,b],X,Y)} succeeds with computed answer 
$\theta = \{\mathtt{Y/[a,b|X]}\}$.
In fact, \texttt{append([a,b],X,[a,b|X])}$\in \mathit{M^{SL}_{\tt
    APPEND}}$, and $\theta$ is simply-local wrt.\ the query above.

On the other hand, we can also say that 
the query \texttt{append(X,[a,b],Y)} has \emph{no successful
input-consuming derivations}.
In fact, for every $A \in \mathit{M^{SL}_{\tt APPEND}}$ we have that
the first input position of $A$ is filled in by a non-variable term.
Therefore there is no simply-local $\theta$ such that
\texttt{append(X,[a,b],Y)}$\theta  \in \mathit{M^{SL}_{\tt APPEND}}$.  This shows that this
semantics allows us to model correctly deadlocking derivations.

However, if one considers derivations that are not input-consuming,
then the query \texttt{append(X,[a,b],Y)}
has successful derivations. Likewise,
if one considers substitution that are not simply-local, 
 \texttt{append(X,[a,b],Y)} has instances in
$\mathit{M^{SL}_{\tt APPEND}}$.
\end{myexample}

\subsection{Modelling Partial Derivations}
\label{subsec:incomplete}

Dynamic scheduling also allows for parallelism. In this
context it is important to be able to model the result of partial
 derivations. 
That is to say, instead of considering computed answer substitutions
for complete derivations, we now consider computed answer
substitutions for partial derivations. 
 As we will see, this will be essential
in order to prove termination of the programs.

Let $\mathit{SM}_P$ be the set of all simply-moded atoms of the
extended Herbrand universe of $P$. In analogy to
Prop.~\ref{propo:least-model-exists}, we have the following proposition.

\begin{propo} 
Let $P$ be simply-moded. Then 
$\mathit{T^{SL}_P}\uparrow
  \omega(\mathit{SM}_P)$ is the least simply-local model of $P$
containing $\mathit{SM}_P$. 
\qed\end{propo}

We denote the least
simply-local model of $P$ containing $\mathit{SM}_P$ by 
$\mathit{PM}^{SL}_P$, for 
\emph{partial model}.

We now proceed in the same way as in the previous subsection: We first
show a special case of the statement that our semantics is correct,
fully abstract and compositional (for partial derivations). Based on
this, we show the theorem itself.

\begin{rmlemma}
\label{lem:atomic-sem-partial}
Let the program 
$P$ and the atom $A$ be simply-moded.
The following statements are equivalent:
\begin{itemize}
\item[(i)]
  there exists an input-consuming partial
 derivation
$A\stackrel{\vartheta}  \Dersym_P~\ol A$,
\item[(ii)]  there exists 
a substitution $\theta$ such that $A\theta\in \mathit{PM}^{SL}_P$ and 
$\mathit{In}(A\theta)=\mathit{In}(A)$,
\end{itemize}
where $A\vartheta$ and $ A\theta$ are variant.

\begin{proof}
The proof is almost identical to the one of Lemma \ref{lem:atomic-sem-total}.
The basic difference is that now, in the base cases, we have to consider
derivations of length zero.

$\mathrm{(i)}\Rightarrow \mathrm{(ii)}.$
If $\mathit{len}(\delta)=0$,
then $\ol A=A$ and $\vartheta=\epsilon$ (the empty substitution).
The thesis follows from the fact that $A$ is simply-moded and 
$\mathit{PM^{SL}_P}$ contains the set of all simply-moded atoms.

$\mathrm{(ii)}\Rightarrow \mathrm{(i)}.$
If $A\theta\in \mathit{T^{SL}_P}\uparrow 0(\mathit{SM}_P)=\mathit{SM}_P$ 
 then $\theta$ is just a renaming of the output variables of $A$. The thesis follows by taking $\vartheta$ to be the empty substitution
and $\delta$ to be the derivation of length zero.
\end{proof}
\end{rmlemma}

We now state that our 
semantics is correct, fully abstract and compositional for partial derivations.

\begin{rmtheorem}
\label{thm:comp-sem-partial}
Let the program $P$ and the query $\ol A$ be simply-moded.  The
following statements are equivalent:
\begin{itemize}
\item[(i)] there exists an input-consuming derivation $\ol
  A\stackrel{\vartheta} \Dersym_P~\ol A'$,
\item[(ii)] there exists a substitution $\theta$,
 simply-local wrt.\ 
  $\ol A$,
 such that $\ol
  A\theta\in \mathit{PM}^{SL}_P$,
\end{itemize}
where $\ol A\theta$ is a variant of  $\ol A\vartheta$.  

\begin{proof}
The proof is analogous to the one of Theorem \ref{thm:comp-sem-total}, 
but using Lemma~\ref{lem:atomic-sem-partial} instead of
\ref{lem:atomic-sem-total}. 
\end{proof}
\end{rmtheorem}

Note that the derivation in point (i) ends in $\ol A'$, which might be
non-empty.

\begin{myexample}
  Consider again \texttt{APPEND}.  First, $\mathit{PM}^{SL}_{\tt
    APPEND}$ contains $\mathit{M^{SL}_{\tt APPEND}}$ as a subset (see
  Ex.\ \ref{complete-model-ex}). Note that $\mathit{M^{SL}_{\tt
      APPEND}}$ is obtained by starting from the fact clause
  \verb:append([],Ys,Ys): and repeatedly applying the
  $\mathit{T^{SL}_P}$ operator using the recursive clause of {\tt
    APPEND}. Now to obtain the remaining atoms in
  $\mathit{PM}^{SL}_{\tt APPEND}$, we must repeatedly apply the
  $\mathit{T^{SL}_P}$ operator, starting from any simply moded atom,
  i.e., an atom of the form ${\tt append}(s,t,x)$ where $s$ and $t$
  are arbitrary terms but $x$ does not occur in $s$ or $t$. It is easy
  to see that we thus have to add $\mathit{SM}_P$ together with
\[
\begin{array}{ll}
\{ 
{\tt append}([t_1,\ldots,t_n|s],t,[t_1,\ldots,t_n|x])\ |\  &
t_1,\ldots,t_n,s,t  \mbox{ are arbitrary terms,}\\
& 
x \mbox{ is a fresh variable}\}.
\end{array}
\]
Consider the query $\mathtt{append([a,b|X],Y,Z)}$.
The substitution $\theta = \{\mathtt{Z/[a,b|Z']}\}$ 
is simply-local wrt.\ the query, and 
$\mathtt{append([a,b|X],Y,[a,b|Z'])}\in \mathit{PM^{SL}_{\tt APPEND}}$.
Using Theorem \ref{thm:comp-sem-partial}, we can conclude that the query
has a partial derivation with computed
answer $\theta$.
Following the same reasoning, one
can also conclude that the query has a partial derivation with
computed answer $\theta = \{\mathtt{Z/[a|Z']}\}$.
\end{myexample}

\section{Termination}\label{sec:termination}
Input-consuming derivations were originally conceived as an abstract
and ``reasonably strong'' assumption about the selection rule in order
to prove termination~\cite{S99}. The first result in this area was a
sufficient criterion applicable to well- and nicely-moded programs.
This was improved upon by dropping the requirement of well-modedness,
which means that one also captures termination by
deadlock~\cite{BER99}. In this section, we only consider {\em simply}
moded programs and queries (simply-moded and  well-moded
  programs form two largely overlapping, but distinct classes),
  and we provide a criterion for termination which is sufficient and
  {\em necessary}, and hence an exact characterisation of termination.
  We first define our notion of termination.

\begin{rmdefinition}
  A program is 
called 
\emph{input terminating} iff all its
  input-consuming derivations started in a simply-moded query are
  finite.
\qed
\end{rmdefinition}

In order to prove that a program is input terminating, we need the
concept of moded level mapping \cite{EBC99}.

\begin{rmdefinition} 
  A function $|\;|$ is a \emph{moded level mapping} iff it
  maps atoms into $\mathbb{N}$ and such that for any $\mathbf{s}$, $\mathbf{t}$ and
  $\mathbf{u}$, $|p(\mathbf{ s},\mathbf{ t})|=|p(\mathbf{ s},\mathbf{
    u})|$. \qed
\end{rmdefinition}

The condition $|p(\mathbf{ s},\mathbf{ t})|=|p(\mathbf{ s},\mathbf{
  u})|$ states that the \emph{level} of an atom is independent from
the terms in its output positions.
%
%

Note that programs without recursion terminate trivially. In this
context, we need the following standard definitions~\cite{Apt97}. 
\begin{rmdefinition}
  Let $P$ be a program, $p$ and $q$ be relations.  We say that 
\begin{itemize}
\item \emph{p refers to q} iff there is a clause in $P$ with
  $p$ in the head and $q$ in the body. 
\item \emph{p depends on q} iff $(p,q)$ is in the reflexive and
  transitive closure of the relation \emph{refers to}.
\item \emph{p} and \emph{q} are \emph{mutually recursive}, written 
$p \simeq q$, iff $p$ and $q$ depend on each other.
\qed
\end{itemize}
\end{rmdefinition}

We now define {\em simply-acceptability}, which is in analogy
to acceptability~\cite{AP94}, but defined to deal with simply-moded and
input-consuming programs.

\begin{rmdefinition}
  Let $P$ be a program and
  $M$ a simply-local model of $P$ containing $\mathit{SM}_P$. 
A clause $H\leftarrow \mathbf{A},B,\mathbf{ C}$ is 
  \emph{simply-acceptable wrt.\ the moded level mapping 
$|\;|$ and $M$} iff for
  every substitution $\theta$ simply-local wrt.~it,
\[
\mbox{if }  \ol A\theta \in M
\mbox{ and } \mathit{Rel}(H)\simeq  \mathit{Rel}(B)
\mbox{ then }
|H\theta|>|B\theta|.
\]
The program $P$ is  \emph{simply-acceptable wrt.\ $M$} iff 
there exists a moded level mapping $|\;|$ such that each clause
of $P$ is simply-acceptable wrt.\ $|\;|$ and $M$.
\end{rmdefinition}

We also say that $P$ is \emph{simply-acceptable} if it is
simply acceptable wrt.~some $M$. 

Let us compare simply-acceptability to acceptability,
used to prove left-termination \cite{AP94}.
Acceptability is based on a
(classical) model $M$ of the program, and for a clause $H \la A_1,\dots,A_n$,
one requires $|H \theta| > |A_i \theta|$ only if 
$M \models (A_1,\dots,A_{i-1}) \theta$. The reason is that for 
LD-derivations, $A_1,\dots,A_{i-1}$ must be completely resolved before $A_i$ is
selected. 
By the correctness of LD resolution~\cite{Apt97}, it turns out that the
c.a.s.\  $\theta$, just before $A_i$ is selected, is such that
$M \models (A_1,\dots,A_{i-1})\theta$.  It has been argued previously
that it is difficult to use a similar argument for input-consuming
derivations~\cite{S99}.  Using the results of the previous section, we
have overcome this problem. We exploited that provided that programs
and queries are simply-moded, we know that even though
$A_1,\dots,A_{i-1}$ may not be resolved completely,
$A_1,\dots,A_{i-1}\theta$ will be in any ``partial model'' of the
program.

In the next two subsections, we prove that simply-acceptability is a
sufficient and necessary criterion for termination. 
The sections can be skipped by the reader who is not interested in the 
proofs. 

\subsection{Sufficiency of Simply-Acceptability}
The following corollary of \cite[Lemma 5.8]{BER99} allows us to
restrict our attention to queries containing only one atom.

\begin{rmcoro}
\label{cor:inf-one-atom}
Let $P$ be a simply-moded program. $P$ is input terminating iff for
each simply-moded one-atom query $A$ all input-consuming derivations
of $P\U\{A\}$ are finite.
\end{rmcoro}

From now on, we say that a relation $p$ is \emph{defined in} the program
$P$ if $p$ occurs in a head of a clause of $P$, and that $P$
\emph{extends} the program $R$ iff no relation defined in $P$ occurs
in $R$.

The following theorem is actually even more general than the one
in~\cite{BERS01}. It shows that simply-acceptability is a sufficient
criterion for termination, and can be used in a modular way.

\begin{rmtheorem}
\label{ter-modular}
Let $P$ and $R$ be two simply-moded programs such that $P$ extends $R$. 
Let $M$ be a simply-local model of $P\cup R$ containing $\mathit{SM}_P$. 
Suppose that 
\begin{itemize}
\item $R$ is input terminating,
\item $P$ is simply acceptable 
wrt.\ $M$ (and a moded level mapping $|\;|$).
\end{itemize}
Then $P\cup R$ is input terminating.

\begin{proof}
 First, for each predicate symbol $p$, we define $\mathit{ dep}_P(p)$ to
  be the number of predicate symbols it depends on: 
$\mathit{ dep}_P(p) = \# \{q|\; q
  \mbox{ is defined in } P \mbox{ and } p\sqsupseteq q\}$.
Clearly,
  $\mathit{ dep}_P(p)$ is always finite. Further, it is immediate to see
  that if $p\simeq q$ then $\mathit{ dep}_P(p)=\mathit{ dep}_P(q)$ and that if
  $p\sqsupset q$ then $\mathit{ dep}_P(p)>\mathit{ dep}_P(q)$.

 We can now prove our theorem. By Corollary \ref{cor:inf-one-atom},
  it is sufficient to prove that for any simply-moded one-atom query
  $A$, all input-consuming derivations of $P\U\{A\}$ are finite.

  First notice that if $A$ is defined in $R$ then the result follows
  immediately from the hypothesis that $R$ is input terminating and
  that $P$ is an extension of $R$. So we can assume that $A$ is
  defined in $P$.

For the purpose of deriving a contradiction, assume 
 $\delta$ is an infinite input-consuming derivation of $(P\U R)
  \U \{A\}$ such that $A$ is defined in $P$. Then
\[\delta:=A\stackrel{\vartheta_1}\Longrightarrow (B_1,\ldots,B_n)\vartheta_1
\stackrel{\vartheta_2} \Longrightarrow \cdots \]
where $c: H\leftarrow B_1,\ldots,B_n $ is the
input clause used in the first derivation step and 
$\vartheta_1=
\mathit{ mgu}(A , H)$.
 Clearly,
$(B_1,\ldots,B_n)\vartheta_1$ has an infinite 
input-consuming derivation in $P\U R$.
By the Left-Switching lemma, 
for some $i\in\{1,\ldots,n\}$ and for some substitution $\vartheta'_2$,
\begin{enumerate}\parentarabici
\item
there exists 
an infinite input-consuming 
derivation of $(P\U R)\U \{ A\}$ of the form
\[A\stackrel{\vartheta_1}\Longrightarrow (B_1,\ldots,B_n)\vartheta_1
\stackrel{\vartheta'_2} \longmapsto \mathbf{ C}, (B_i,\ldots,B_n)
\vartheta_1\vartheta'_2\cdots ;\]
\item there exists an infinite input-consuming derivation
of $P\U \{B_i\vartheta_1\vartheta'_2\}.$
\end{enumerate}

Let  $\theta=(\vartheta_1\vartheta'_2)_{|c}$.
By Lemma \ref{lemma:ic-cas}, $\theta$ is simply-local
wrt.\ $c$.
 Consider the instance $H\theta\leftarrow
(B_1,\ldots,B_n)\theta $ of $c$.
By Theorem \ref{thm:comp-sem-partial},
$(B_1,\ldots,B_{i-1})\theta\in M$.

We  show that (2) cannot hold, by induction on
$\langle \mathit{ dep}_P(\mathit{ Rel}(A)), |A|\rangle$ with respect
to the ordering $\succ$ defined by: $\langle m,n\rangle \succ \langle
m',n'\rangle$ iff either $m>m'$ or $m=m'$ and $n>n'$.

{\it Base}.  Let $\mathit{ dep}_P(\mathit{Rel}(A))=0$ ($|A|$ is arbitrary).
 In
this case, $A$ does not depend on any predicate symbol of $P$, thus
all the $B_i$ as well as all the atoms occurring in its descendents in
any input-consuming derivation are defined in $R$. The hypothesis that
$R$ is input terminating contradicts $(2)$ above.

\emph{Induction step}. 
We distinguish two cases:
\begin{enumerate}\parentalphi 
\item $\mathit{ Rel}(H)\sqsupset \mathit{ Rel}(B_i)$,
\item $\mathit{ Rel}(H)\simeq \mathit{ Rel}(B_i)$.
\end{enumerate}
In case $(a)$ we have that 
$
\mathit{ dep}_P(\mathit{ Rel}(A))
=\mathit{ dep}_P(\mathit{ Rel}(H\theta))>
\mathit{ dep}_P(\mathit{ Rel}(B_i\theta)).
$. 
Therefore, 
$$
\langle \mathit{ dep}_P(\mathit{ Rel}(A)), |A|\rangle=
\langle \mathit{ dep}_P(\mathit{ Rel}(H\theta)), |H\theta|\rangle \succ
\langle \mathit{ dep}_P(\mathit{ Rel}(B_i\theta)), |B_i\theta|\rangle.
$$
In case $(b)$,
from the hypothesis
 that $P$ is simply-acceptable wrt.\ $|\;|$ and $M$, $\theta$ is simply-local wrt.\ $c$ 
 and $(B_1,\ldots,B_{i-1})\theta\in M$, it follows that
$|H\theta|> |B_i\theta|$.
Consider the partial input-consuming derivation $A\stackrel{\theta}
\longmapsto \mathbf{ C}, (B_i,\ldots,B_n) \theta $. By Lemma
\ref{lem:prop-in*}
 and the fact that $|\;|$ is a moded level mapping, we
have that $|A|=|A\theta|=|H\theta|$.  Hence, $\langle \mathit{
  dep}_P(\mathit{ Rel}(A)), |A|\rangle =\langle \mathit{
  dep}_P(\mathit{ Rel}(H\theta)), |H\theta|\rangle \succ \langle
\mathit{ dep}_P(\mathit{ Rel}(B_i\theta)), |B_i\theta|\rangle$.  In
both cases, the contradiction follows by the inductive hypothesis.
\end{proof}
\end{rmtheorem}

The above theorem suggests proving termination in a modular way, i.e.,
extending a program that is already known to be input-terminating by
a program that is simply-acceptable. Of course, this theorem holds in
particular if the former program is empty.

\begin{rmtheorem}
\label{ter1}
Let $P$ be a simply-moded program.
If $P$ is simply-acceptable then it is input
  terminating.

\begin{proof}
The proof follows from Theorem \ref{ter-modular}, by setting $R=\emptyset$.
\end{proof}
\end{rmtheorem}

\subsection{Necessity of Simply-Acceptability}

We now prove the converse of Theorem \ref{ter1}. The results of the
previous and this subsection together provide an exact
characterisation of input termination.

\begin{rmdefinition}
\label{def:i-tree}
An \textit{IC-tree} for $P\cup\{Q\}$ 
via a dynamic selection rule ${\cal R}$ is a tree such that
\begin{itemize}
\item its root is $Q$,
\item every node $Q'$ with a selected atom $A$ has exactly one descendant
$Q''$ for each clause $c$ such that $Q''$ is an input-consuming resolvent
of $Q'$ wrt.\ $A$ and $c$.
\end{itemize}
\end{rmdefinition}

Notice that, since we are considering dynamic selection rules,
it can happen that there is no atom selectable (i.e., meeting the condition
about input-consuming resolvents above) in a node of
an IC-tree.

\begin{rmdefinition}
An \emph{LIC-derivation} is an input-consuming derivation in which at each step
the selected atom is the leftmost atom which can be resolved via
an input-consuming derivation step.
\end{rmdefinition}

Similarly, we define the notion of LIC-tree.

\begin{rmdefinition}
An \emph{LIC-tree} is an IC-tree 
in which at each node
the selected atom is the leftmost atom which can be resolved via
an input-consuming derivation step.
\end{rmdefinition}

Branches of LIC-trees are LIC-derivations.

\begin{rmlemma}\textbf{([LIC-tree 1)}\ 
\label{lemma:lic-tree1}
An LIC-tree for $P\cup\{Q\}$ is finite iff all LIC-consuming
derivations of
$P\cup\{Q\}$ are finite.

\begin{proof}
By definition, the LIC-trees are finitely branching.
The claim now follows by K\"{o}nig's Lemma.
\end{proof}
\end{rmlemma}

For a program $P$ and a query $Q$, we denote by
 $\mathit{lnodes}^\mathit{ic}_P(Q)$
the number of nodes in an LIC-tree for $P\cup\{Q\}$.
The following property of IC-trees will be needed.

\begin{rmlemma}
\label{lemma:lnodes}
Let the program $P$ and the query $\ol A,B$ be simply-moded.
Suppose that $P$ is input terminating and $\ol A\theta\in
\mathit{PM}^{SL}_P$, where $\theta$ is a simply-local substitution 
wrt.~$\ol A$. Then
$\mathit{lnodes}^{\mathit{ic}}_P(\ol A,B)\geq 
\mathit{lnodes}^{\mathit{ic}}_P(B\theta)$.

\begin{proof}
Consider an LIC-tree $T$ for $P\cup\{\ol A,B\}$.
By the hypothesis that $\ol A\theta\in
\mathit{\mathit{PM}^{SL}_P}$, it follows that there exists
a substitution
$\sigma$ (possibly the empty substitution) such that
$\theta$ is more general than $\sigma$, and
$\sigma$ is a partial c.a.s.
for $P\cup \{\ol A\}$,
 such that either no atom is selectable 
or $B\sigma$ is the selected atom.

Clearly, $\mathit{lnodes}^{\mathit{ic}}_P(\ol A,B)\geq 
\mathit{lnodes}^{\mathit{ic}}_P(B\sigma)$.
By Lemma~\ref{input-pushing-lemma} we have $\mathit{lnodes}^{\mathit{ic}}_P(B\sigma)\geq
\mathit{lnodes}^{\mathit{ic}}_P(B\theta)$. Hence the thesis.
\end{proof}
\end{rmlemma}

We are now in position to prove necessity of simply-acceptability.
\begin{rmtheorem}
\label{ter2}
Let $P$ be a simply-moded program.
If $P$ is 
input terminating then  $P$ is simply-acceptable.

\begin{proof}
We show that there exists a moded level mapping $|\;|$ for $P$
 such that
$P$ is simply-acceptable  wrt.\ $|\;|$ and $\mathit{PM^{SL}_P}$,
the latter being the least 
simply-local model of $P$  containing $\mathit{SM}_P$.

Given an atom $A$, we denote with $A^*$ an atom obtained from $A$
by replacing the terms filling in its output positions with fresh
distinct variables. Clearly, we have that $A^*$ is simply-moded.
Then we define the following moded level mapping 
for $P$:
\[ |A| = \mathit{lnodes}^\mathit{ic}_P(A^*). \]
Notice that, the level $|A|$
of an atom $A$ is independent from the terms filling in its output 
 positions, i.e., $|\;|$ is a moded level mapping. Moreover, since $P$ is 
input terminating and $A^*$ is simply-moded,
 all the input-consuming derivations of 
$P\cup\{A^*\}$ are finite.
Therefore, 
by Lemma~\ref{lemma:lic-tree1}, 
$\mathit{lnodes}^\mathit{ic}_P(A^*)$  is defined (and finite), and thus
$|A|$ is defined (and finite) for every atom $A$.

We now prove that $P$ is simply-acceptable wrt.\  $|\;|$ and 
$\mathit{PM^{SL}_P}$.

Let  $c: H\leftarrow \mathbf{ A},B,\mathbf{ C}$ be a clause of $P$ and 
$ H\theta\leftarrow \mathbf{ A}\theta,B\theta,\mathbf{ C}\theta$
be an instance of $c$ where 
 $\theta$ is a simply-local substitution  wrt.\ $c$.
We show that 
\[
\mbox{if } \mathit{PM^{SL}_P} \models \ol A\theta 
\mbox{ and } \mathit{Rel}(H)\simeq  \mathit{Rel}(B)
\mbox{ then }
|H\theta|>|B\theta|.
\]

Consider a variant
 $c': H'\leftarrow \mathbf{ A}',B',\mathbf{ C}'$ of $c$ variable disjoint from 
 $(H\theta)^*$. 
Let $\rho$ be a renaming such that $c'=c\rho$.
Clearly,
 $(H\theta)^*$ and $H'$ unify.
Let $\mu=\mathit{mgu}((H\theta)^*, H')=\mathit{mgu}((H\theta)^*, H\rho)$.
Since $\mu$ is simply-local wrt
$(H\theta)^*$ and $H'$, we have 
$\mathit{Dom}(\mu)\subseteq \mathit{Var}(\mathit{Out}((H\theta)^*))\cup
\mathit{Var}(\mathit{In}(H\rho))$.  Hence
 $(\mathbf{ A}',B',\mathbf{ C}')\mu=
(\mathbf{ A},B,\mathbf{ C})\rho\mu$,
and
$$(H\theta)^*\stackrel{\mu}\Longrightarrow (\mathbf{ A},B,\mathbf{ C})\rho\mu$$
is an input-consuming derivation step, i.e.,
$(\mathbf{ A},B,\mathbf{ C})\rho\mu$
is a descendant of $(H\theta)^*$ in an LIC-tree for $P\cup\{(H\theta)^*\}$.

Moreover,
$(\mathbf{ A},B,\mathbf{ C})\rho\mu\approx
(\mathbf{ A},B,\mathbf{ C})(\rho\mu)_{|\mathit{In}(H)}=
(\mathbf{ A},B,\mathbf{ C})\theta_{|\mathit{In}(H)}$.

Let
$\theta=\theta_{|\mathit{In}(H)}\theta_{|\mathit{Out}(\ol A)}\theta_{|\mathit{Out}(B,\ol C)}$.
Hence, $\theta_{|\mathit{Out}(\ol A)}$ is simply-local wrt.\ 
$A\theta_{|\mathit{In}(H)}$.
Then, we have that
\[
\begin{array}{lclll}
|H\theta| & = &  \mathit{lnodes}^\mathit{ic}_P((H\theta)^*) & (\mbox{by definition of $ |\;|$})\\
& > &  \mathit{lnodes}^\mathit{ic}_P((\mathbf{ A},B,\mathbf{ C})\theta_{|\mathit{In}(H)}) & (\mbox{by definition of LIC-tree})\\
& \geq &  \mathit{lnodes}^\mathit{ic}_P((\mathbf{ A},B)\theta_{|\mathit{In}(H)}) & (\mbox{by definition of LIC-tree})\\
& \geq & \mathit{lnodes}^\mathit{ic}_P((B\theta_{|\mathit{In}(H)}\theta_{|\mathit{Out}(\ol A)}) & (\mbox{by Lemma \ref{lemma:lnodes}})\\
& = & \mathit{lnodes}^\mathit{ic}_P((B\theta)^*) & (\mbox{since $\theta$ is simply-local wrt.\ c})\\
 & = & |B\theta|  & (\mbox{by definition of $ |\;|$}).
\end{array}
\]
\end{proof}
\end{rmtheorem}

\subsection{A Characterisation}
Summarising, we have characterised input termination by simply-acceptability.

\begin{rmtheorem}
\label{ter}
A simply-moded program $P$ is simply-acceptable iff it is input
 terminating.
In particular,  if $P$ is input terminating, then it is
simply-acceptable wrt.\ $\mathit{PM^{SL}_P}$.

\begin{proof}
By Theorem \ref{ter1} and Theorem \ref{ter2}. 
\end{proof}
\end{rmtheorem}

\begin{figure}[t]
\begin{program}
\% \> quicksort(Xs, Ys) \la Ys {\rm is an ordered permutation of } Xs.
\\[2mm]
 \>         quicksort(Xs,Ys) \la quicksort\_dl(Xs,Ys,[]).\\[2mm]
 \>         quicksort\_dl([X|Xs],Ys,Zs)  \la partition(Xs,X,Littles,Bigs),\\
 \> \>         quicksort\_dl(Bigs,Ys1,Zs). \\ 
 \> \>         quicksort\_dl(Littles,Ys,[X|Ys1]),\\ 
 \>         quicksort\_dl([],Xs,Xs).\\[2mm]
 \>         partition([X|Xs],Y,[X|Ls],Bs) \la  X =< Y, partition(Xs,Y,Ls,Bs).\\
 \>         partition([X|Xs],Y,Ls,[X|Bs]) \la X > Y,  partition(Xs,Y,Ls,Bs).\\
 \>         partition([],Y,[],[]).\\[2mm]

\> mode  quicksort(In,Out).\\
\> mode  quicksort\_dl(In,Out,In).\\
\> mode  partition(In,In,Out,Out).\\
\> mode  =<(In,In).\\
\> mode  >(In,In).
\end{program}
\caption{The \texttt{quicksort} program\label{fig:qs}}
\end{figure}

\begin{myexample}
Figure~\ref{fig:qs} shows program 15.3 from \cite{SS86}: \texttt{quicksort}
using a form of difference lists (we permuted two body atoms for the
sake of clarity). This program is simply-moded, and when 
used in combination with dynamic scheduling, the standard delay
declarations for it are the following:
\begin{program}
  \> delay  quicksort(Xs, \_) until nonvar(Xs).\\
  \> delay  quicksort\_dl(Xs, \_, \_) until nonvar(Xs).\\
  \> delay  partition(Xs, \_, \_, \_) until nonvar(Xs).\\
  \> delay  =<(X,Y) until ground(X) and ground(Y).\\
  \> delay >(X,Y) until ground(X) and ground.(Y)
\end{program}
The last two declarations fall out of the scope of Lemma
\ref{lem:delay=IC}. Nevertheless, if we think of the built-ins
\texttt{>} and \texttt{=<} as being conceptually defined by a program
containing infinitely many ground facts of the form
\texttt{>($n$,$m$)}, with $n$ and $m$ being two appropriate integers,
the derivations respecting the above delay declarations are exactly
the input-consuming ones.  We can prove that the program is input
terminating. Define \emph{len} as
\[
\begin{array}{rclll}
len([h|t]) & = & 1 + len(t),\\
len(a)     &=& 0 &          \mbox{ if $a$ is not of the form $[h|t]$}.
\end{array}
\]
We use the following moded level mapping  (positions with $\_$
are irrelevant)
\begin{eqnarray*}
|\mathtt{quicksort\_dl}(l,\_,\_)| &=& len(l),\\
|\mathtt{partition}(l,\_,\_,\_)| &=& len(l).
\end{eqnarray*}
The level mapping of all other atoms 
can be set to 0. Concerning the model, the simplest solution is to use 
the model that expresses the dependency between the list lengths of
the arguments of $\mathtt{partition}$, i.e., $M$ should
contain all atoms of the form
$\mathtt{partition}(l_1,x,l_2,l_3)$ where
$len(l_1) > len(l_2)$ and
$len(l_1) > len(l_3)$.
\end{myexample}

\section{Benchmarks}\label{app:benchmarks}
\label{sec:benchmarks}
\begin{sloppypar}
In order to assess how realistic the conditions of Lemma
\ref{lem:delay=IC} are, we have looked into three collections of
logic programs, and we have checked whether those programs were simply
moded (\textbf{SM}), input-consistent (\textbf{IC}) and whether they
satisfied both sides of Lemma \ref{lem:delay=IC} (\textbf{L}).
Notice that  programs which are not input-consistent do not satisfy
the conditions of Lemma \ref{lem:delay=IC}.  For
this reason, some \textbf{L} columns are left blank.
%
%
The results, reported in Tables \ref{1} to \ref{3}, show that our results apply to
the majority of the programs considered.  Table~\ref{1} shows the
programs from Apt's collection \cite{Apt97,AP94}, Table~\ref{2} those of
the DPPD collection 
(\texttt{http://dsse.ecs.soton.ac.uk/$\sim$mal/systems/dppd.html}), and 
Table \ref{3} some programs from Lindenstrauss's collection
(\texttt{http://www.cs.huji.ac.il/$\sim$naomil}).
\end{sloppypar}


\newcommand{\ns}{\hspace{-0.15em}}

\begin{table}[t]
\begin{center}
{\footnotesize
\begin{tabular}{|>{\ns\tt}l|>{\ns\rm}c|>{\ns\rm}c|>{\ns\rm}c||>{\ns\tt}l|>{\ns\rm}c|>{\ns\rm}c|>{\ns\rm}c|}
\hline
 & {\bf SM} & {\bf IC} &   {\bf  L} & &  {\bf SM} & {\bf IC} &   {\bf  L}\\
\hline
append(In,In,Out) & yes  & yes & yes &   
mergesort(Out,In) & no &  & \\
\hline
append(Out,Out,In) & yes  & yes &  no & 
mergesort\_variant(In,Out,In) & yes & yes & no\\
\hline
append3(In,In,In,Out) & yes & yes & yes & 
ordered(In) & yes & no &\\
\hline
color\_map(In,Out) & yes & no &  &  
overlap(In,In) & yes & no & \\
\hline
color\_map(Out,In) & yes & yes &  yes &  
overlap(In,Out) & yes & yes &  yes\\
\hline
dcsolve(In,\_) & yes & yes & yes  & 
overlap(Out,In) & yes & yes & yes\\
\hline
even(In) & yes & no &  & 
perm\_select(In,Out) &  yes & yes & no\\
\hline
fold(In,In,Out) & yes & yes & yes & 
perm\_select(Out,In) &  yes & yes & no\\
\hline
list(In) & yes & yes & yes & 
qsort(In,Out) & yes & yes & yes\\
\hline
 lte(In,In) & yes & yes & no & 
qsort(Out,In) & no &  &\\
\hline
 lte(In,Out) & yes & yes & yes & 
reverse(In,Out) & yes & yes & yes\\
\hline
 lte(Out,In) & yes & yes & no & 
reverse(Out,In) & yes & yes & yes \\
\hline
map(In,In) & yes & yes & yes &  
select(In,In,Out) & yes & no &\\
\hline
map(In,Out) & yes & yes & yes & 
select(Out,In,Out) & yes & yes & yes\\
\hline
map(Out,In) & yes & yes & yes & 
subset(In,In) & yes & no &\\
\hline
member(In,In) & yes & no &  & 
subset (Out,In) & yes & yes & yes\\
\hline
member(In,Out) & yes & yes & yes & 
sum(In,In,Out) & yes & yes & yes\\
\hline
member(Out,In) & yes & yes & yes &  
sum(Out,Out,In) & yes & yes & yes \\
\hline
mergesort(In,Out) & yes & no & &
 type(In,In,Out) & no & & \\
\hline
\end{tabular}}
\end{center}
\caption{Programs from Apt's Collection\label{1}}
\end{table}
%

\begin{table}[t]
\begin{center}
{\footnotesize
\begin{tabular}{|>{\ns\tt}l|>{\ns\rm}c|>{\ns\rm}c|>{\ns\rm}c||>{\ns\tt}l|>{\ns\rm}c|>{\ns\rm}c|>{\ns\rm}c|}
\hline
 & {\bf SM} & {\bf IC} &   {\bf  L} & & {\bf SM} & {\bf IC} &   {\bf  L}\\
\hline
applast(In,In,Out) & yes  & yes & yes &   relative (In,Out) & yes & yes & yes\\
\hline
depth(In,Out) & yes & no &  & relative (Out,In) & yes & yes & yes \\
\hline
flipflip(In,Out) & yes & yes & yes  & rev\_acc(In,In,Out) & yes & yes & yes\\
\hline
flipflip(Out,In) & yes & yes & yes &  rotate(In,Out)  & yes & yes & yes\\
\hline
generate(In,In,Out) & yes & no & &rotate(Out,In)  & yes & yes & yes \\
\hline
liftsolve(In,In) & yes & yes & yes & solve(In,In,Out) ) & yes & no &\\
\hline
liftsolve(In,Out) & yes & yes & yes &  square\_square(In,Out) & yes & yes & yes \\
\hline
 match(In,In) & yes & no & &  squretr(In,Out)  & yes & yes & yes\\
\hline
 match\_app(In,In) & yes & yes & no& ssupply(In,In,Out) & yes & yes & yes \\
\hline
  match\_app(In,Out) & yes & yes & no & trace(In,In,Out)   & yes & no &\\
\hline
 max\_lenth(In,Out,Out) & yes & yes & yes & trace(In,Out,Out)   & no &  & \\
\hline
memo\_solve(In,Out) & yes & no & &transpose(In,Out) & yes & no &\\
\hline
 prune(In,Out) & yes & no & &transpose(Out,In) & yes & yes & yes\\
\hline
 prune(Out,In) & yes & no & &unify(In,In,Out) & yes & no & \\
\hline
\end{tabular}}
\end{center}
\caption{Programs from  DPPD's Collection\label{2}}
\end{table}

\begin{table}[t]
\begin{center}
{\footnotesize
\begin{tabular}{|>{\ns\tt}l|>{\ns\rm}c|>{\ns\rm}c|>{\ns\rm}c||>{\ns\tt}l|>{\ns\rm}c|>{\ns\rm}c|>{\ns\rm}c|}
\hline
 & {\bf SM} & {\bf IC} &   {\bf  L} & &  {\bf SM} & {\bf IC} &   {\bf  L} \\
\hline
ack(In,In,\_) & yes & yes & no & huffman(In,Out) &  no &  &\\
\hline
concatenate(In,In,Out) & yes & yes & yes & huffman(In,Out) &  no &  &\\
\hline
credit(In,Out) & yes & yes & yes & normal\_form(\_,In) & yes & no & \\
\hline
deep(In,Out) & yes & yes & yes & queens(In,Out) & yes & yes & yes\\
\hline
deep(Out,In)  & no &  & &queens(Out,In) & yes & yes & no\\
\hline
descendant(In,Out) & yes & yes & yes  & rewrite(In,Out) & yes & no & \\
\hline
descendant(Out,In) & yes & yes & yes & transform(In,In,In,Out) & yes & yes & yes  \\
\hline
holds(In,Out) & yes & yes & yes  &twoleast(In,Out) & no &  &\\
\hline
\end{tabular}}
\end{center}
\caption{Programs from  Lindenstrauss's Collection\label{3}}
\end{table}

\section{Conclusion}
\label{sec:comparison}
In this paper, we have proven a result that \emph{demonstrates} --- for
a large class of programs --- the equivalence between 
delay declarations and input-consuming derivations. This was only
speculated in \cite{BER99,BER00-cl2000}.  In fact, even though the
class of programs we are considering here (simply-moded programs) is
only slightly smaller than the one of nicely-moded programs considered
in \cite{BER99,BER00-cl2000}, for the latter a result such as Lemma
\ref{lem:delay=IC} does not hold.
  
We have provided a denotational semantics for input-consuming
derivations using a variant of the well-known $T_P$-operator. 
Our semantics follows the  
\mbox{$s$-semantics} approach~\cite{BGLM94} and
thus enjoys the typical properties of semantics in this class. This
semantics improves on the one introduced in \cite{BER00-cl2000} in two
respects: The semantics of this paper models (within a uniform
framework) both complete and incomplete derivations, and there is no
requirement that the program must be well-moded. 

Falaschi {\em et al.}~\cite{FGMP97} have defined a denotational
semantics for CLP programs with dynamic scheduling of a somewhat
different kind: the semantics of a query is given by a set of closure
operators; each operator is a function modelling a possible effect of
resolving the query on a program state (i.e., constraint on the program
variables). However, we believe that our approach is more suited to 
termination proofs.

As mentioned in Subsec.~\ref{subsec:incomplete}, in the context of 
parallelism and concurrency \cite{Nai88}, one can have derivations
that never {\em succeed}, and yet compute substitutions. Moreover,
input-consuming derivations essentially correspond to the execution
mechanism of (Moded) FGHC~\cite{UM94}. Thus we have provided a
model-theoretic semantics for such programs/programming languages,
which go beyond the usual success-based SLD resolution mechanism of
logic programming. 

On a more practical level, our semantics for partial derivations is
used in order to prove termination.  We have provided a necessary and
sufficient criterion for termination, applicable to a wide class of
programs, namely the class of simply-moded programs. For instance, we
can now prove the termination of \texttt{QUICKSORT}, which is not
possible with the tools of
\cite{BER99,S99} (which provided only a sufficient condition).
In the termination proofs, we exploit that any selected atom in an
input-consuming derivation is in a model for partial derivations, in a
similar way as this is done for proving left-termination. It is only
on the basis of the semantics that we could present a characterisation
of input-consuming termination for simply-moded programs.

\bibliographystyle{plain} 
\bibliography{jan,sandro}

\begin{thebibliography}{10}

\bibitem{Apt90}
K.~R. Apt.
\newblock Introduction to {L}ogic {P}rogramming.
\newblock In J.~van Leeuwen, editor, {\em Handbook of Theoretical Computer
  Science}, volume B: Formal Models and Semantics, pages 495--574. Elsevier,
  Amsterdam and The MIT Press, Cambridge, 1990.

\bibitem{Apt97}
K.~R. Apt.
\newblock {\em From Logic Programming to Prolog}.
\newblock Prentice Hall, 1997.

\bibitem{AE93}
K.~R. Apt and S.~Etalle.
\newblock On the unification free {P}rolog programs.
\newblock In A.~Borzyszkowski and S.~Sokolowski, editors, {\em Proceedings of
  MFCS '93}, LNCS, pages 1--19. Springer-Verlag, 1993.

\bibitem{AL95}
K.~R. Apt and I.~Luitjes.
\newblock Verification of logic programs with delay declarations.
\newblock In V.~S. Alagar and M.~Nivat, editors, {\em Proceedings of AMAST'95},
  LNCS, pages 66--90. Springer-Verlag, 1995.
\newblock Invited Lecture.

\bibitem{AP94}
K.~R. Apt and D.~Pedreschi.
\newblock Modular termination proofs for logic and pure {P}rolog programs.
\newblock In G.~Levi, editor, {\em Advances in Logic Programming Theory}, pages
  183--229. Oxford University Press, 1994.

\bibitem{BER99}
A.~Bossi, S.~Etalle, and S.~Rossi.
\newblock Properties of input-consuming derivations.
\newblock {\em ENTCS}, 30(1), 1999.
\newblock http://www.elsevier.nl/locate/entcs.

\bibitem{BER00-cl2000}
A.~Bossi, S.~Etalle, and S.~Rossi.
\newblock Semantics of input-consuming programs.
\newblock In J.~Lloyd, editor, {\em CL 2000}, number 1048 in LNCS, pages
  33--45. Springer-Verlag, 2000.

\bibitem{BERS01}
A.~Bossi, S.~Etalle, S.~Rossi, and J.-G. Smaus.
\newblock Semantics and termination of simply-moded logic programs with dynamic
  scheduling.
\newblock In D.~Sands, editor, {\em Proceedings of the European Symposium on
  Programming}, LNCS. Springer-Verlag, 2001.

\bibitem{BGLM94}
A.~Bossi, M.~Gabbrielli, G.~Levi, and M.~Martelli.
\newblock The $s$-semantics approach: theory and applications.
\newblock {\em Journal of Logic Programming}, 19/20:149--197, 1994.

\bibitem{EBC99}
S.~Etalle, A.~Bossi, and N.~Cocco.
\newblock Termination of well-moded programs.
\newblock {\em Journal of Logic Programming}, 38(2):243--257, 1999.

\bibitem{FGMP97}
M.~Falaschi, M.~Gabbrielli, K.~Marriott, and C.~Palamidessi.
\newblock Constraint logic programming with dynamic scheduling: {A} semantics
  based on closure operators.
\newblock {\em Information and Computation}, 137:41--67, 1997.

\bibitem{HL94}
P.~M. Hill and J.~W. Lloyd.
\newblock {\em The {G}{\"o}del Programming Language}.
\newblock The MIT Press, 1994.

\bibitem{sicstus98}
Intelligent Systems Laboratory, Swedish Institute of Computer Science, PO Box
  1263, \mbox{S-164 29} Kista, Sweden.
\newblock {\em {SICS}tus {P}rolog {U}ser's Manual}, 1998.
\newblock {\tt http://www.sics.se/sicstus/docs/3.7.1/html/sicstus\_toc.html}.

\bibitem{K79}
R.~A. Kowalski.
\newblock {A}lgorithm = {L}ogic + {C}ontrol.
\newblock {\em Communications of the ACM}, 22(7):424--436, 1979.

\bibitem{Llo87}
J.~W. Lloyd.
\newblock {\em Foundations of Logic Programming}.
\newblock Symbolic Computation -- Artificial Intelligence. Springer-Verlag,
  1987.

\bibitem{Na86}
L.~Naish.
\newblock {\em Negation and {C}ontrol in {P}rolog}, volume 238 of {\em LNCS}.
\newblock Springer-Verlag, 1986.

\bibitem{Nai88}
L.~Naish.
\newblock Parallelizing {NU}-{P}rolog.
\newblock In R.~A. Kowalski and K.~A. Bowen, editors, {\em Proceedings of
  ICLP/SLP '88}, pages 1546--1564. MIT Press, 1988.

\bibitem{Smaus-thesis}
J.-G. Smaus.
\newblock {\em Modes and Types in Logic Programming}.
\newblock PhD thesis, University of Kent at Canterbury, 1999.
\newblock Available from {\tt http://www.cs.ukc.ac.uk/pubs/1999/986/}.

\bibitem{S99}
J.-G. Smaus.
\newblock Proving termination of input-consuming logic programs.
\newblock In D.~De Schreye, editor, {\em Proceedings of ICLP'99}, pages
  335--349. MIT Press, 1999.

\bibitem{SS86}
L.~Sterling and E.~Shapiro.
\newblock {\em The Art of Prolog}.
\newblock MIT Press, 1986.

\bibitem{UM94}
K.~Ueda and M.~Morita.
\newblock {M}oded {F}lat {GHC} and its message-oriented implementation
  technique.
\newblock {\em New Generation Computing}, 13(1):3--43, 1994.

\end{thebibliography}

\end{document}